\pdfoutput=1
\documentclass[preprint,10pt]{elsarticle}

\usepackage{amsmath, amsthm, amssymb,amssymb,graphicx,fancyhdr,fancyhdr,helvet,ifpdf,setspace,graphics}
\usepackage{booktabs}
\usepackage{pstricks}
\usepackage{cancel}
\usepackage{fullpage}
\usepackage{enumerate}
\usepackage[normalsize]{subfigure}
\usepackage[section]{placeins}
\journal{Journal Name}

\begin{document}
\begin{frontmatter}
\title{Characterization of Coupled Turbulent Wind-wave Flows Based on Large Eddy Simulation}

 \author[lsu]{Tianqi Ma}
 \author[lsu]{Chao Sun\corref{cor}}
 \ead{csun@lsu.edu}
 \cortext[cor]{Corresponding author}
 \address[lsu]{Department of Civil and Environmental Engineering, Louisiana State University, Baton Rouge, Louisiana 70803, USA}

%%%%%%%%%%%%%%%%%%%%%%%%%%abstract%%%%%%%%%%%%%

\begin{abstract}

Wind-wave interaction involves wind forcing on wave surface and wave effects on the turbulent wind structures, which essentially influences the wind and wave loading on structures. Existing research on wind-wave interaction modeling ignores the inherent strong turbulences of wind. The present study aims to characterize the turbulent airflow over wave surfaces and wave dynamics under wind driving force. A high-fidelity two-phase model is developed to simulate highly turbulent wind-wave fields. Instead of using uniform wind, inherent wind turbulences are prescribed at the inlet boundary using the turbulent spot method. The developed model is validated by comparing the simulated wind-wave flow characteristics with experimental data. With the validated model, a numerical case study is conducted on a $10^2$ m scale under extreme wind and wave conditions. The result shows that when inherent wind turbulences are considered, the resultant turbulence is strengthened and is the summation of the inherent turbulence and the wave-induced turbulence. In addition, the wave coherent velocities and shelter effect are enhanced because of the presence of wind inherent turbulence. The regions of intense turbulence depend on the relative speed between wind velocity and wave phase speed. Higher wind velocities induce greater turbulence intensities, which can be increased by up to 17\%. Different relative speed between wind and wave can induce opposite positive-negative patterns of wave coherent velocities. The wave-coherent velocity is approximately proportional to the wind velocity, and the influenced region mainly depends on the wave heights.
\end{abstract}


\end{frontmatter}

\section{Introduction}

\label{S:1}

Strong tropical cyclones always incur extreme winds, surge, and waves that cause extensive damage to critical infrastructure of coastal communities. During such high category hurricanes, the wind and wave are coupled through exchanging momentum flux and heat flux at the air-water interface. The wind wave interaction involves wind forcing on the wave surface and the wave influence on the turbulent air structures and pressure distributions. As a result, the wind-wave interaction changes the wind-wave flow field characteristics and influences the combined wind-wave pressures on structures. Hence, it is crucial to understand the evolution of turbulent air flow over wave surface and wave dynamics under wind driving force. Wave shape (wave steepness) and wave age are two critical parameters to describe momentum and energy between air and water. Momentum is transferred from the atmosphere to the ocean during the wave development. The ratio between the phase speed ($C_{p}$) of the peak wave and the wind speed at 10 m height ($U_{10}$) or wind friction velocity ($u*$) is defined as the wave age ($\frac{C_{p}}{U_{10}}$ or $\frac{C_{p}}{U*}$). Based on the wave age, the wave regime is divided into two types, the wind waves and the swells. The wind waves move more slowly than the wind and the waves are wind-driven waves. The swells move faster than the wind, and the wind is considered as wave-driven wind \cite{sullivan2010dynamics,sullivan2014large}.

%basic theory

The understanding of generation and growth of wind-driven waves is developed based on macroscopic experiments and relevant theories. There are two primary mechanisms, one is a laminar instability mechanism and the other is an aerodynamic sheltering mechanism, to describe the wind-driven wave generation and development. The laminar instability mechanism originating from the Kelvin-Helmholtz shear instability theory \cite{lamb1924hydrodynamics}, is based on stability analysis of the coupled air-water system. 
%the gravity and surface tension
Later on, Miles \cite{miles1957generation} proposed an elaborate instability theory to explain the generation and growth of waves generated by wind. The existing waves caused shear instability associated with critical layer which in turn forced the waves to grow. Phillips \cite{phillips1966dynamics} found that the resonance between waves and air pressure fluctuations leaded to early-stage wave generation. Phillips \cite{miles1978dynamics} later combined his turbulent pressure generation mechanism with Miles’ critical wave growth mechanism. The aerodynamic sheltering mechanism was derived from the pressure differential of airflow past a blunt body. Jeffrey \cite{jeffreys1925formation} proposed that the pressure differential was caused by airflow separation and induced the wave growth . Motzfeld \cite{stanton1932growth} found that the pressure differential caused by sheltering effect was too weak to explain the growth rate of waves. Belcher and Hunt \cite{belcher1998turbulent} pointed out that the wave-coherent turbulence forces also caused pressure differential and induced wave growth. 

To better understand the wind wave interaction, extensive experimental studies have been conducted in the past decades.
%field experiments
Dobson \cite{dobson1971measurements} measured wave elevation and atmospheric pressure on wind-driven sea waves through field testing. Results suggested that rates of energy transferred from wind to wave is an order of magnitude larger than the theoretical value predicted by Miles. Further experimental studies (\cite{snyder1981array}, \cite{hasselmann1991field}) showed that the Miles' model overestimates the energy transferring rate especially for low frequency waves, and the error is within an order of magnitude. Recently, Grare \textit{et al.} \cite{grare2013wave} measured turbulence velocity profiles above waves and correlated them to the phase of waves. It was found that Miles' critical layer mechanism is valid for a certain range of wave ages. However, the near-surface velocity field and turbulence can not be captured due to the technical limitations of measuring equipment.
%lab experiments
Kharif \textit{et al.} \cite{kharif2008influence} analyzed the following wind effects on freak waves through laboratory testing and numerical modeling. Results indicated that winds shift the focus point and increase the peak wave amplitude. Tian and Choi \cite{tian2013evolution} experimentally studied the following wind effect on dispersive focusing waves. It was concluded that the Miles' model predicts well for weak wind, but its performance declines for strong wind. 
%PIV method
Furthermore, Particle Image Velocimetry (PIV) techniques are used to directly measure water-side or air-side velocity field, tangential viscous stress, instantaneous surface vorticity and viscous stress. Banner and Peirson \cite{banner1998tangential} measured near-surface tangential stresses. The authors found that prior to the formation of wind waves, the tangential stress contributed to the entire wind stress. However, the wave form drag dominated the wind stress after the wave developed. Veron \textit{et al.} \cite{veron2007measurements} directly measured the tangential viscous stress along the wind-wave interface and instantaneous surface vorticity. The authors observed the airflow separation over young wind waves. Buckley \textit{et al.} \cite{buckley2016structure} investigated the structure of the turbulent airflow above water surface in a laboratory using a imaging system. The authors found that the averaged mean and turbulent quantities are strongly phase-locked. The structure of the airflow above wave surface, momentum fluxes, and turbulence production regions, significantly depend on the wave age.

%numerical simulations
In addition to experimental studies, numerical methods have been applied to analyze the coupling effects between wind and wave. In the past decades, wind influences on waves have been studied extensively based on one-phase and two-phase models. In the one-phase models, the wave was treated as a wavy surface moving at a specific speed and the airflow motion is only considered above the wave surface \cite{de1997direct}. Large Eddy Simulation (LES) was applied to study the instantaneous airflow over a wide range of wave ages \cite{sullivan2008large}. However, the wave induced wind fields were simulated over wavy boundary conditions without considering the wind effects on waves. The wind-wave interaction was not coupled directly \cite{zou1995viscoelastic,sullivan2000simulation,sullivan2008large}. As a result, one-phase models can only capture the motion of either the air or the water. In comparison, two-phase flow models, which are more physics-based, consider two-way dynamic wind-wave coupling. A two-way coupling direct numerical simulation (DNS) approach can be used to model the air and water dynamics simultaneously. Yang and Shen \cite{yang2010direct} presented a multi-domain DNS approach to simulate stress-driven turbulent Couette flows over wave surfaces to study turbulence in the vicinity of water waves. The Naiver-Stokes equation is solved in the air and water domain separately. Two domains were coupled by enforcing velocity continuity and balance of stress at the explicit water-air interface. Yet this method is unsuitable for cases that have complex interfaces with large deformation and fragmentation\cite{zou2017wind}. To resolve the problem in the multi-domain DNS approach, a single-domain DNS approach was proposed where the water-air interface is always modeled by the volume of fluid (VOF) surface capturing technique. The single domain method overcomes the difficulties in modeling complex interfaces and can simulate violent dynamic events such as breaking waves. Iafrati \textit{et al.} \cite{iafrati2013modulational} presented a 2D DNS method to study the deep-water wave breaking induced by modulational instability without external wind and its contribution to the air-water interaction. The authors found that wave breaking rises large-scale dipole structures in the air phase, which dissipate most of the energy induced by the breaking wave. Since DNS methods are computationally intensive and hardly affordable, RANS (Reynolds-averaged Navier–Stokes) and LES turbulence models are frequently used to solve the Navier-Stokes equations. RANS-VOF models were used to study the wind effects on wave overtopping in \cite{hieu2014study} and breaking solitary waves in \cite{xie2014numerical}. Due to lack of testing data with wind presence, these two models were only validated for cases without wind. Zou and Chen \cite{zou2017wind} used a Navier-Stokes solver in OpenFOAM with a subgrid-scale-stress (SGS) LES model and VOF air-water interface capturing technique to study the wind and current effects on wave formation and breaking. The authors found that the winds shift the wave focus point mainly due to the action of wind-driven current instead of direct wind forcing. The vertical shear of the wind-driven current plays an important role in wave propagation and transformation.

In summary, existing studies on wind-wave interaction are mainly based on small scale laboratory testing and focus on the air-water interface because of the limitation of the measurement techniques. The one phase flow model treats wave as a wavy surface moving at a specific speed and doesn't consider the interaction between wind and wave. Although the two-phase model couples the wind and wave, research efforts have been primarily exerted on the evolution of wave profiles under wind forcing. In addition, existing research on wind-wave interaction modeling focuses mainly on uniform wind-driven effects on formation, evolution, and breaking of waves. Turbulences of wind are only simulated near the air-water interface. The inherent strong turbulences of wind above the air-water interface and the vertical turbulence transport are missing, which are inadequate to accurately quantify the combined wind-wave pressures on structures exposed to extreme wind-wave flows simultaneously.

To fill this knowledge gap, the present study aims to characterize the turbulent air fields above air-water surfaces and their coupling effect. A high-fidelity two-phase flow model is developed to simulate the highly turbulent coupled wind-wave flow fields. The developed wind wave solver is validated with experimental results. The validated solver is then applied to analyze the flow field characteristics of extreme wind and wave on a $10^2$ m scale. The reminder of the paper is as follows. A two-phase model for highly turbulent wind-wave flow field is developed in Section 2. The post process method to analyze the complex wind structures above waves is presented in Section 3. Simulated wind and wave fields are verified through comparing the elevations of waves and phase averaged wind structure with experimental data in Section 4. The results of simulated wind and wave fields under extreme conditions are discussed in Section 5. Conclusions are summarized in Section 6.

\section{Model Description}
\label{S:2}

\subsection{Governing Equations}
The water and air are assumed as incompressible flow. The mass conservation and Navier-Stokes equations are

\begin{equation}
\label{eq:11}
\nabla \cdot \mathbf{U} = 0
\end{equation}

\begin{equation}
\label{eq:22}
 \frac{\partial{\rho \mathbf{U}}}{\partial{t}} + \nabla \cdot (\rho \mathbf{UU}) = -\nabla p_{d} - \mathbf{g}\cdot\mathbf{x}\nabla{\rho} +\nabla \cdot (\mu_{eff}\nabla \mathbf{U})+ \nabla{\mathbf{U}}\cdot\nabla{\mu_{eff}} + \sigma\kappa\nabla{\alpha}
\end{equation}
where $\mathbf{U}$ is the velocity vector; $\rho$ is the fluid density; $p_{d}=p-\rho\mathbf{g}\cdot \mathbf{x}$ is the dynamic pressure by subtracting the hydrostatic part from total pressure $p$; $\mathbf{g}$ is the gravitational acceleration; $\mathbf{x}$ is the position vector; $\mu_{eff}=\rho(\nu+\nu_t)$ is the effective dynamic viscosity, in which $\nu$ and $\nu_t$ are the kinematic and turbulent eddy viscosity respectively. $\nu_t$ is modeled by a turbulence model. $\sigma\kappa\nabla{\alpha}$ is a surface tension term where $\sigma$ is the surface tension coefficient, $\kappa$ is the free surface curvature, and $\alpha$ is the volume fraction. In this study, $\sigma$ is set as 0.07 N/m. The volume of fluid (VOF) method is used to mark the air-water interface. An indicator scalar function $\alpha = \frac{V_{water}}{V_{cell}}$ is used to represent the fractional volume of a cell occupied by water. For a two-phase air-water flow, $\alpha=1$ represents that the cell is full of water, $\alpha=0$ represents the cell is full of air, and $0<\alpha<1$ indicates that the cell contains the free surface. Then the density $\rho$ and viscosity $\mu$ of fluid at an arbitrary location can be written as
\begin{equation}
\label{eq:33}
\rho = \alpha\rho_{water}+(1-\alpha)\rho_{air}
\end{equation}

\begin{equation}
\label{eq:44}
\mu = \alpha\mu_{water}+(1-\alpha)\mu_{air}
\end{equation}
where $\rho_{water}$ and $\rho_{air}$ represent water density and air density respectively; $\mu_{water}$ and $\mu_{air}$ are water viscosity and air viscosity. With adopting the VOF method, the two immiscible fluids (water and air) are modeled as an effective continuous flow and are governed by one set of conservation equations. The advection of the indicator function is governed by the transport equation:

\begin{equation}
\label{eq:55}
\frac{\partial\alpha}{\partial t}+\nabla\cdot(\mathbf{U}\alpha)+\nabla\cdot[\mathbf{U_r}\alpha(1-\alpha)]=0
\end{equation}
where a compression term $\nabla\cdot[\mathbf{U_r}\alpha(1-\alpha)]$ is introduced to reduce the interface smearing effect. The compression term is acting only in the free surface zone because of the inclusion of $(1-\alpha)\alpha$. $\mathbf{U_r}$ is a compression velocity.

The turbulent kinetic viscosity $\nu_t$ is modeled by a turbulence model. In this study, the large eddy simulation (LES) method is adopted with large scale eddies resolved and small scale turbulences modeled with the subgrid-scale stress (SGS) model. A wall adapting local eddy viscosity (WALE) is adopted to model the subgrid-scale stress, which can model accurate wall boundary layer\cite{nicoud1999subgrid,ben2017assessment,weickert2010investigation}. The WALE model calculates the eddy viscosity based on the invariants of the velocity gradients.

\begin{equation}
\label{eq:66}
\nu_t = (C_w\Delta)^2\frac{(S_{ij}^{d}S_{ij}^d)^{3/2}}{(\bar S_{ij}\bar S_{ij})^{5/2}+(S_{ij}^d S_{ij}^d)^{5/4}}
\end{equation}
where $\bar S_{ij}=1/2(\partial \bar u_i/\partial x_j+\partial \bar u_j/\partial x_i)$ is the rate-of-strain tensor; $\bar{( )}$ represents filtered variable; parameter $S_{ij}^d$ is the traceless symmetric part of the square of the velocity gradient tensor, which is calculated using Eqn. (\ref{eq:77}).

\begin{equation}
\label{eq:77}
S_{ij}^d = \frac{1}{2}\left(\frac{\partial \bar u_k}{\partial x_i}\frac{\partial \bar u_j}{\partial x_k}+\frac{\partial \bar u_k}{\partial x_j}\frac{\partial \bar u_i}{\partial x_k}\right)-\frac{1}{3}\delta_{ij}\frac{\partial \bar u_k}{\partial x_l}\frac{\partial \bar u_l}{\partial x_k}
\end{equation}
where $\delta _{ij}$ is the Kronecker’s delta.

\subsection{Boundary conditions}
\subsubsection{Wave wind boundary conditions at inlet boundary}
In the present study, the wave is generated at the inlet boundary through specifying the free surface elevation and water particle velocities, which is implemented referring to the olaFlow framework developed by Higuera et al.\cite{olaFlow}. Different waves, linear wave and non-linear waves, can be generated at the boundary using the classic wave theories. Similarly, the wind velocity is prescribed for the air phase at the inlet boundary. Uniform or other (log or power) mean wind profiles can be generated above the water surface, where the wind velocity in the air-water interface grid is consistent with the water velocity at wave surface. In addition to a mean velocity, the wind field at the inlet is prescribed as the summation of the mean flow and turbulent fluctuations. To generate the turbulent wind, a turbulent spot method is implemented to generate spatially and temporally correlated turbulent fluctuations, which possesses prescribed Reynolds stress and integral length and satisfies continuity constraint \cite{kroger2018generation,poletto2013new}. In turbulent spots method, a set of turbulent spots are randomly distributed at the inlet boundary and are convected by the mean velocity through the boundary. For the $i^{th}$ spot, an inner velocity distribution is set as $u_n^i(\mathbf{r}-\mathbf{r}^i)=\varepsilon_{n}^i f_n(\mathbf{r}-\mathbf{r}^i)$, where n is the component number, $\mathbf{r}^i$ is the center of the spot, $\varepsilon_{n}^i$ is a uniformly distributed random number between -1 and 1. The inner velocity distribution determines spectra and the integral length scales. The velocity fluctuations at point $\mathbf{r}$ are the sum of contributions from all spots $\mathbf{u}=\sum_{i=1}^N{\mathbf{u}^i(\mathbf{r}-\mathbf{r}^i)}$. This method can generate anisotropic turbulence through introducing the anistropy into the turbulent spots \cite{kornev2007synthesis}. The velocities at the inlet boundary is fixed as specified values with wave particle velocities applied for the water phase and wind velocities for the air phase. The pressure condition at the inlet boundary is set as zero gradient.

\subsubsection{Wave absorption}
A wave relaxation zone is added near the outlet boundary to avoid reflection of waves from outlet boundaries. Base on \cite{jacobsen2012wave}, an explicit relaxation method, $\phi=(1-w_R)\phi_{target}+w_R\phi_{computed}$, is applied to correct the indicator scalar function $\alpha$ and velocity $\mathbf{U}$, where $\phi$ represents $\alpha$ or $\mathbf{U}$, $w_R$ is a weighting function of local coordinate system in the relaxation region. Referring to \cite{fuhrman2006numerical}, an exponential weighted distribution is selected. In this study, the relaxation technique is only effective for waves and has no effect on air flow through setting the weight function as zero in the air phase. The outlet boundary is treated as pressure boundary conditions with the total pressure condition set as fixed values and the velocity condition set as zero gradient.  

\subsection{Numerical Method}

In the present study, the simulations are performed using open-source library OpenFOAM (Open-source Field Operations And Manipulations)\cite{weller1998tensorial}. A standard solver interFoam is provided in OpenFOAM for incompressible two fluid flows, which is based on the Finite Volume Method and VOF surface capturing method \cite{habchi2013partitioned,holzmann2016mathematics}.

\section{Post-process}
\label{S:3}

To analyze the wind-wave coupling interaction, the wave elevation and wind-wave velocity field are sampled during the CFD simulation process. 

\subsection{Wave elevation}
To obtain the surface elevation, several wave gauges are defined along the numerical wave tank. At each gauge location, a linear distribution of points are defined in the vertical direction to obtain the value of $\alpha$ through linear interpolation of the simulated results at the cell center. The water elevation is calculated using the sampled $\alpha$ values:

\begin{equation}
\label{eq:88}
\eta = \sum_{i=1}^{N-1}{(h_{i+1}-h_i)\frac{\alpha_{i+1}+\alpha_{i}}{2}}+h_{min}
\end{equation}
where $h$ is the height of the sampled points; $h_{min}$ is the minimum height of the points along the vertical line.

\subsection{Coordinate transformation and phase detection}
\label{S:3_2}
In the numerical simulation, the wind wave velocities are obtained in the Cartesian coordinate, with $x$, $y$ and $z$ being the streamwise, spanwise and vertical coordinates. The velocity obtained from the simulation has three components, $u$, $v$ and $w$, which are function of ($x$, $y$, $z$). To analyze the air flow above the wave surface, a wave surface following coordinate is introduced. The wind-wave fields are transformed to a orthogonal co-ordinate system as proposed in Ref. \cite{benjamin1959shearing}. The wave is a combination of a series of Fourier wave components. The wave elevation ($\eta$) can be expressed as:
\begin{equation}
\label{eq:1}
\eta \left( x,t \right) =\sum_n{a_ne^{i\left(k_nx-2\omega_nt+\phi _n \right)}}
\end{equation}
where $a_{n}$, $\omega_{n}$, $k_{n}$, $\phi_{n}$ are the amplitude, circular frequency, wave number, and phase of the $n^{th}$ mode. A coordinate system which follows the wave surface is introduced, where the orthogonal co-ordinate($\xi$, $\zeta$) is represented by Cartesian coordinates ($x$, $z$) as:
\begin{eqnarray}
\label{eq:2}
\xi \left( x,z \right) &=& x-i\sum_n{a_ne^{i\left( k_n\xi -2\omega_nt+\phi _n \right)}e^{-k_n\zeta}} \nonumber\\
\zeta \left( x,z \right) &=& z-\sum_n{a_ne^{i\left( k_n\xi -2\omega_nt+\phi _n \right)}e^{-k_n\zeta}}
\end{eqnarray}

In the orthogonal coordinate, the wave surface corresponds to $\zeta =0$. Far away from the surface, the orthogoal coordinate tends to the Cartesian coordinate system. For air turbulence statistics above wave surface, a phase average approach is used to quantify the statistical properties of turbulence in order to study the interaction between wind and surface waves. With wave elevation $\eta$ calculated, wave phase of the wind-wave fields can be achieved by applying a Hilbert transform to the wave profiles. A wave-phase decomposition is applied to the wind-wave velocity field to analyze the air structure above the wave surface. A quantity $q$ can be represented as the sum of a phase-averaged quantity $<q>$ and a turbulent quantity $q'$.
\begin{equation}
\label{eq:4}
q(x,y,z,t)=<q>(\xi,\zeta)+q'(x,y,z,t)
\end{equation}

With wave phase calculated using Hilbert transform to wave profiles, the phase averaged quantity ($<q>(\xi,\zeta)$) can be obtained and the turbulent quantity $q'$ can be obtained by subtracting $<q>$ from $q$. The turbulent quantity $q'$ is used to represent the turbulence of wind or wave velocities. Also, the phase-averaged quantity $<q>$ can be further decomposed into a phase-independent total averaged quantity $\bar{q}$ across all phases and a wave-coherent quantity $\tilde{q}$. Via this decomposing method, the instantaneous 3D turbulence flow field $q(x,y,z,t)$ can be expressed as:

\begin{equation}
\label{eq:5}
q(x,y,z,t)=\bar{q}(\zeta)+\tilde{q}(\xi,\zeta)+q'(x,y,z,t)
\end{equation}

\section{Numerical model validation}
\label{S:4}

To verify the numerical model, experimental scale wave tank models are developed and verified with laboratory testing data. The evolution of wave under wind effect and wind structure above wave surface are selected for comparison between experimental and numerical results. This section separately compares the wave elevation results using experimental data from Ref. \cite{tian2013evolution} and compares the airflow structure using measurements from Ref. \cite{buckley2016structure}.

\subsection{Wind forcing effects on wave profiles}
\label{S:4.1}
To analyze the wind effect on the evolution of waves, dispersive focusing (DF) wave groups were adopted in Ref. \cite{tian2013evolution}. In the present study, a 2D numerical wave tank is developed. The evolution of breaking and nonbreaking dispersive focusing wave groups with and without wind forcing are validated with the experimental data. In the experiment, the wave tank was 15 m long and 1.5 m wide. The water depth was 0.54 m. A piston-type wavemaker was used to generate water waves. A twin-fan blower was used to generate following wind. The ceiling panel of the air passage above the still water surface was 0.45 m. The dispersive focusing wave groups are combined by different wave components.

\begin{equation}
\label{eq:6}
\eta \left( x,t \right) =\sum_{n=1}^N{a_n \cos(k_n x-\omega_{n}t-\phi_{n})}
\end{equation}
where $a_n$ is the amplitude of the $n^{th}$ wave component; $\omega_{n}=2\pi f_{n}$ is the angular frequency. The wave group consists of 128 frequency components. The frequency $f^{n}$ ranges from 1.0 Hz to 2.4 Hz and frequency bandwidth $\Delta f$ is 1.4 Hz. $k_n$ is the wave number, which is obtained by the linear dispersion relation $\omega_n^2=gk_n \tanh(k_n d)$. For each wave component, the wave steepness, $\varepsilon_{n}=k_{n}a_{n}$ was kept constant.  $x$ is the horizontal downstream distance from the position of the wavemaker. The phase $\phi_{n}$ is determined so that the wave groups focus at specified time $t_{b}$ and location $x_{b}$ ($\phi_{n}=k_{n}x_{b}-\omega_{n}t_{b}+2\pi m$). Two wave groups were tested, as shown in Table \ref{table:waveGroupParameters}. The DF 1 wave group remains non-breaking under all wind forcing conditions and DF 2 is a breaking wave group under no wind condition. $f_c$ is the center wave frequency and $f_p$ is the spectral peak frequency. During the experiment, the surface elevation was measured through high-speed imaging. The wind velocities were measured with a anemometer. Temporal surface elevations at four wave gauge locations, which were 2.84m, 5.13m, 7.04 m, and 9.07m downstream from the wave maker, were provided in Ref. \cite{tian2013evolution}. Vertical mean wind velocities at three fetches ($x_f$=1.87 m, 4.87 m and 7.87 m) were calculated.

\begin{table}[h]
\centering
\caption{Specified parameters for the wave group}
\label{table:waveGroupParameters}
\begin{tabular}{l l l l l l l l l l}
\hline
\textbf{Wave group} & \textbf{$f_{c}(Hz)$}&\textbf{$f_{p}(Hz)$}&\textbf{$\Delta f/f_{c}$}&\textbf{$\varepsilon=Nk_{n}a_{n}$}\\
\hline
DF 1 & 1.7 & 1.1 & 0.824 & 0.25 \\
DF 2 & 1.7 & 1.1 & 0.824 & 0.57\\
\hline
\end{tabular}
\end{table}

A 2D numerical wave tank is developed to verify with the experiment. To obtain identical wave elevation as that measured in the experiment, the inlet of the numerical wave tank is set at the location of the first wave gauge ($x$=2.84 m). The surface elevation measured at G1 (Gauge 1) ($x$=2.84 m) is used to drive the wave at inlet boundary in the numerical wave tank. At the inlet, the air flow is fully developed and follows the logarithmic law in the experiment. The wind mean speed generated at the inlet boundary has a log profile in the numerical wave tank. The outlet is defined at $x$=15.0 m (end of the physical wave tank). The height of the computational domain is the same as that of the physical tank, which is 0.99 m. To match the surface elevation measured at G1, the simulated time history of surface elevation is reconstructed with $N = 128$ linear wave components, as implemented in Ref. \cite{zou2017wind}.

\begin{equation}
\label{eq:7}
\eta \left( x,t \right) =\eta _m+\sum_{n=1}^N{a_n\cos \left[ \omega _nt-k_n\left( x-x_m \right) +\phi _n \right]}
\end{equation}
where $\eta_m$ is the mean surface elevation; $x_m$ is the location of the first wave gauge; $a_n$, $\omega_n$ and $\phi_n$ are obtained via fast Fourier transform with respect to the temporal surface elevation at G1. Based on the wave theory, the water particle velocity components are

\begin{eqnarray}
\label{eq:8}
u_w\left( x,z,t \right) &=&\sum_{n=1}^N{a_n\omega _n\frac{\cosh \left[ k_n\left( z+d \right) \right]}{\sinh k_nd}\times \cos \left[ \omega _nt-k_n\left( x-x_m \right) +\phi _n \right]}\nonumber\\
w_w\left( x,z,t \right) &=&\sum_{n=1}^N{-a_n\omega _n\frac{\sinh \left[ k_n\left( z+d \right) \right]}{\sinh k_nd}\times \sin \left[ \omega _nt-k_n\left( x-x_m \right) +\phi _n \right]}
\end{eqnarray}
where $d$ is the water depth. The wind speed at inlet is set with a log profile as that measured in the experiment at $x=4.87 m$. 

\begin{equation}
\label{eq:9}
\frac{u(z)}{u^*}=\frac{1}{\kappa}\ln{\frac{z}{z_0}}
\end{equation}
where $\kappa$ is the von Karman constant and $\kappa =0.41$; $z_0$ is the roughness length; $u^*$ is the friction velocity, which can be estimated through the measured vertical profile. In the numerical model, in order not to distort the wave profiles, the wind velocities within the bottom two layer grids above the water surface are set as the friction velocity. An relaxation zone with a length of 4 m, which is larger than twice the longest wave length, was defined at the end of the numerical wave tank to dissipate the wave energy to eliminate reflective waves. The bottom and top boundaries are set as no-slip walls with zero pressure gradient. A uniform mesh with a grid size of 0.25 cm is generated. The simulation is conducted with an automatic time step with the maximum Courant number set as 0.5. 

\begin{figure}[h!]
\centering
\subfigure[DF 1]{\label{fig:eta_DF1}\includegraphics[width=0.8\linewidth]{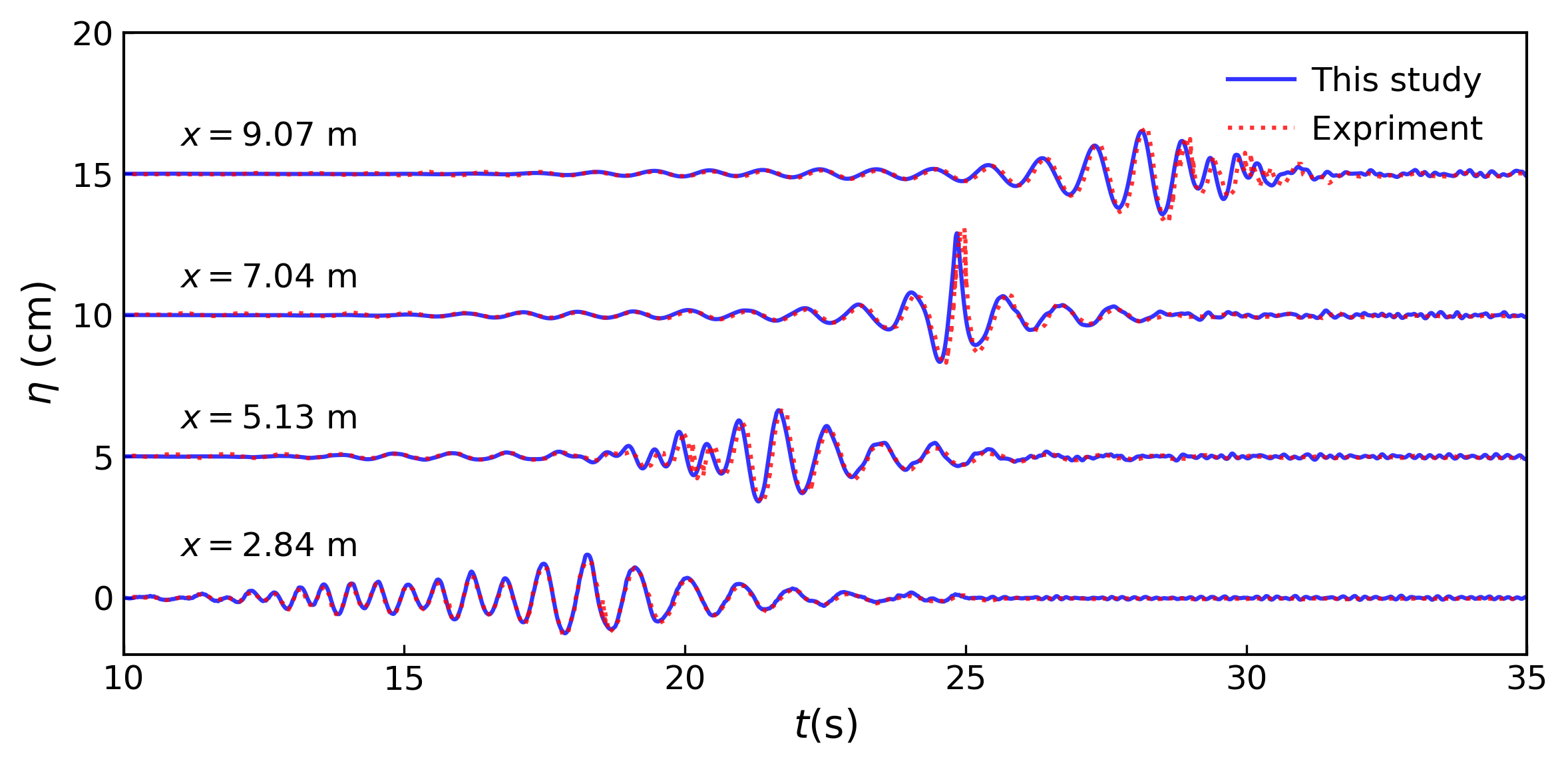}}
\subfigure[DF 2]{\label{fig:eta_DF2}\includegraphics[width=0.8\linewidth]{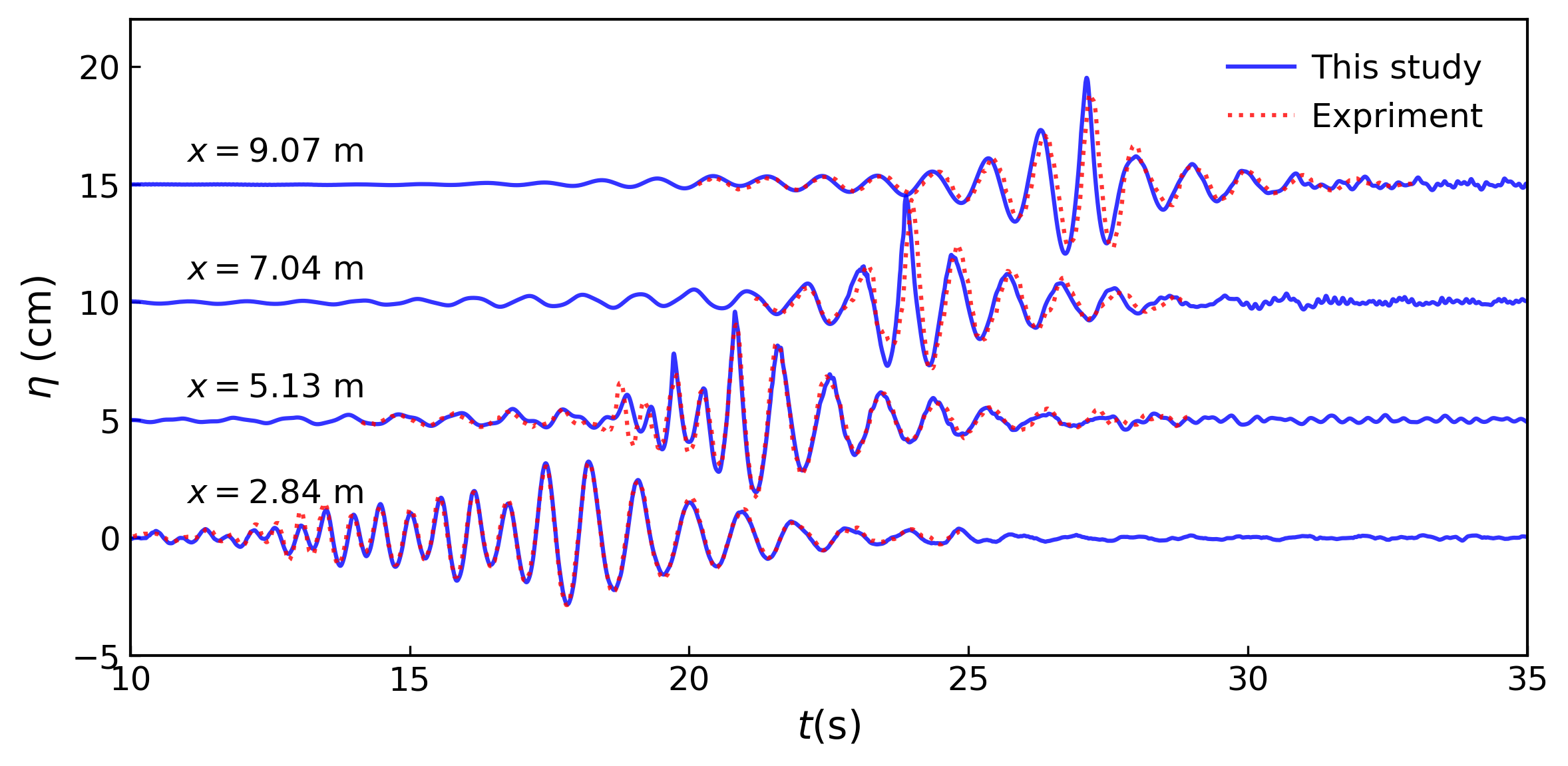}}
\caption{Time series of surface elevation measured at the four gauges without wind forcing. (a) Non-breaking wave group DF1. (b) Breaking wave group DF 2. Blue solid lines denote the results obtained from the numerical model and red dot lines denote the experimental results. The surface elevations at the four gauges are clarified through adding increments of 5 cm, 10 cm and 15 cm to the ordinate.}
\label{fig:waveElevationAtGauges}
\end{figure}

\begin{figure}[h!]
\centering
\includegraphics[width=0.8\linewidth]{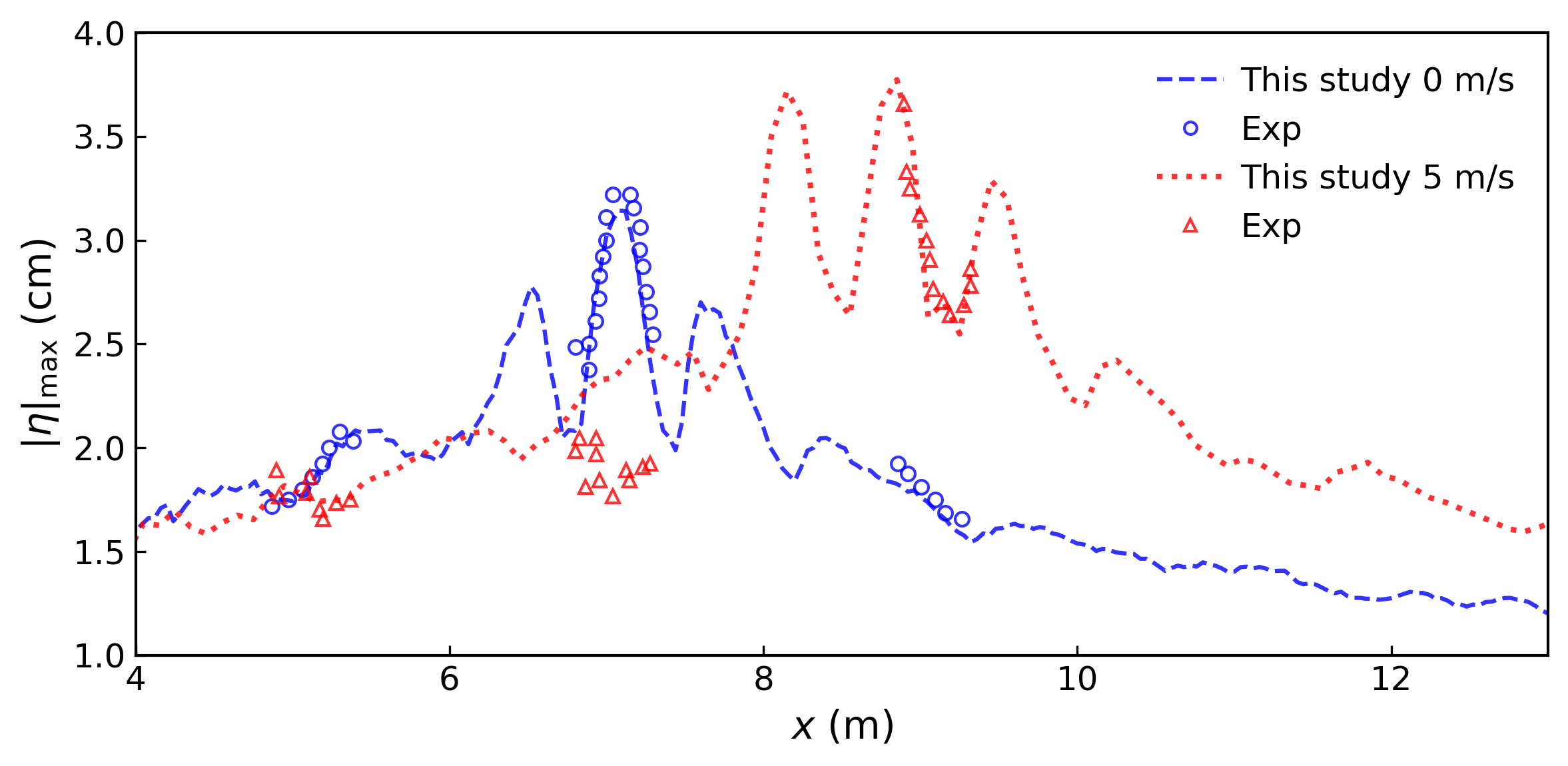}
\caption{Spatial distribution of maximum wave elevation of wave group DF1 under no wind force and following wind force (U = 5 m/s) effect. Blue dashed line and red dot line denote the results obtained from the present numerical model without and with wind forcing. Blue circles and red triangles denote the experimental results obtained by Tian and Choi \cite{tian2013evolution}.}
\label{fig:DF1MaxWaveElevation}
\end{figure}

\begin{figure}[h!]
\centering
\includegraphics[width=0.8\linewidth]{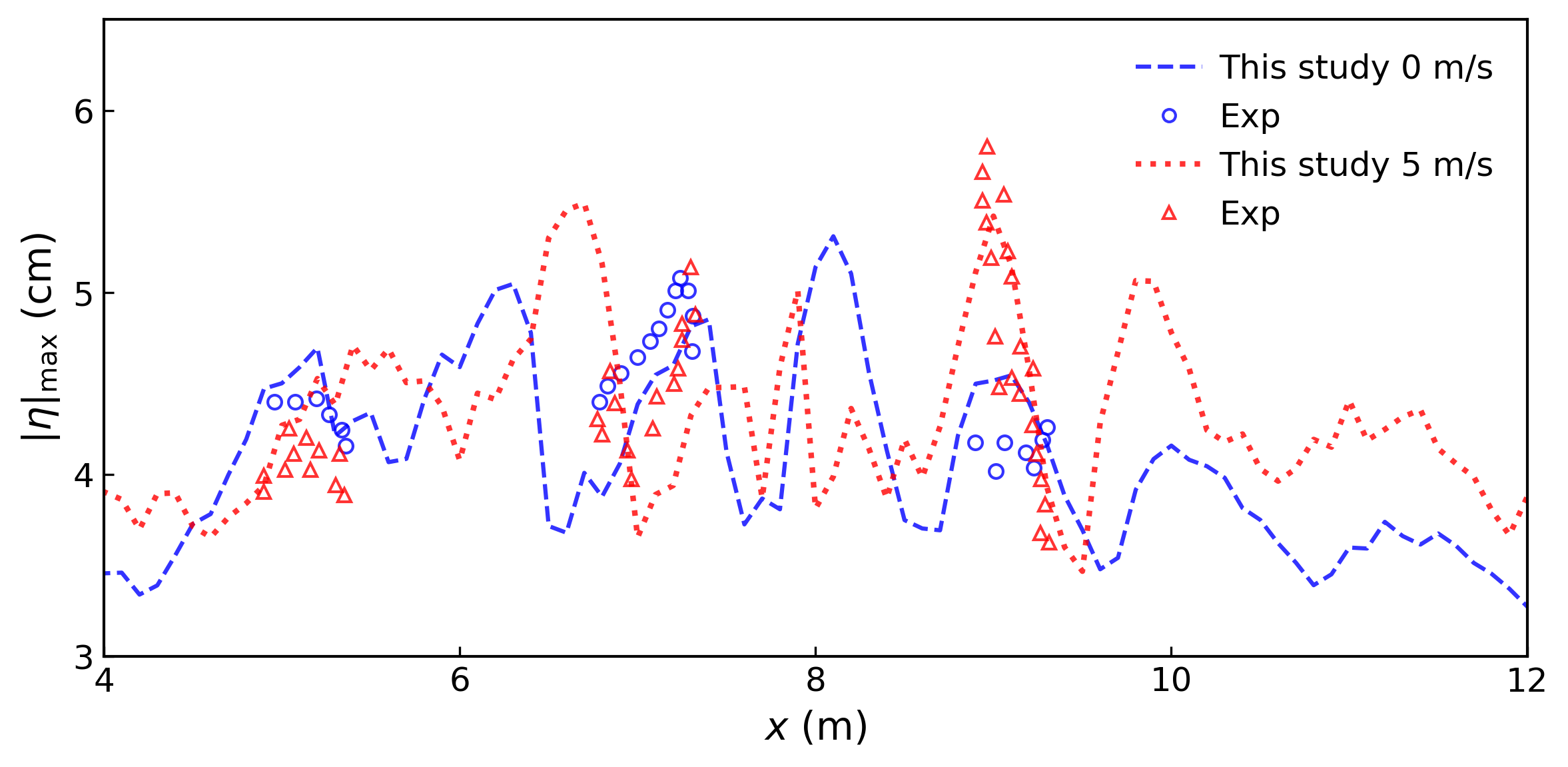}
\caption{Spatial distribution of maximum wave elevation of wave group DF2 under no wind force and following wind force (U = 5 m/s) effect. Blue dashed line and red dot line denote the results obtained from the numerical model without and with wind forcing. Blue circles and red triangles denote the experimental results obtained by Tian and Choi \cite{tian2013evolution}.}
\label{fig:DF2MaxWaveElevation}
\end{figure}

The simulate evolution of wave groups without wind and under following wind is compared with the experimental data in Tian and Choi \cite{tian2013evolution}. Fig. \ref{fig:waveElevationAtGauges}(a) and (b) show the comparisons between the simulated and measured surface elevation at the four wave gauges without wind forcing for non-breaking and breaking wave groups. It shows that the simulated surface elevations agree well with the experimental results without wind forcing. Fig. \ref{fig:DF1MaxWaveElevation} shows the comparisons between the simulated and measured spatial distribution of maximum wave elevations for the DF1 wave group. The spatial distributions of maximum wave elevation under no wind forcing matches well with the measurements for the non-breaking wave group. Under $U=5$ m/s, the numerical model predicts accurate maximum surface elevation at the focusing points where the maximum surface elevation occurs. Comparison of the spatial distribution of the maximum surface elevation between cases without wind forcing and with wind forcing shows that the following wind causes the downstream shift of focus point and increase of the peak surface elevation at the focus point. The numerical maximum surface elevations of breaking wave DF2 are compared with measurement in Fig. \ref{fig:DF2MaxWaveElevation}. In Fig. \ref{fig:DF2MaxWaveElevation}, the simulated maximum wave heights agree with the experimental results yet slight underestimation can be observed near the breaking point (x is around 7 m) because finer meshes are needed to accurately capture the interface when breaking happens.

\subsection{Characteristics evolution of wind field above air-water interface}
\label{S:4.2}

Numerical results of wind forcing effect on wave elevation profile are verified with experimental results in the previous section. This section verifies the numerical model via comparing numerical and experimental results of the wind field characteristics above the wave surface. In different wave regimes, the air-sea momentum flux is different. To compare with experimental results in Ref. \cite{buckley2016structure}, two representative wind wave conditions are selected, with wave ages ($C_{p}/u*$) being 3.7 and 31.7, as list in Table \ref{table:CaseSet}. The friction velocity ($u_{*}$) and the velocity extrapolated at a height of 10 m $U_{10}$ were calculated by fitting the logarithmic part of the mean air velocity profile. The peak wave frequency $f_{p}$ was obtained from WG (wave gauge) frequency spectra. The peak amplitude $a_{p}$ was computed from the wave elevations. The wave speed $C_{p}$, the wave length $\lambda_{p}$, and wave number $k_{p}$ were derived by applying linear wave theory to $f_{p}$. Cases for young wind wave and wind over swells are simulated in this section. In the experiment, the wave tank was 42 m long, 1 m wide, and 1.25 m high. The water depth was 0.7 m. For wind over swell cases, the regular waves were generated by a mechanical wave maker. For young wind waves, the waves were generated by wind. To avoid reflection of waves, a wave absorbing beach was placed at the end of the tank to dissipate wave energy. A recirculating wind tunnel was used to generate wind. Using a high resolution imaging system, the two dimensional airflow above the wave surface was measured at a fetch of 22.7m downstream from the wind entry. 

\begin{table}[h]
\centering
\caption{Summary of experimental conditions. The parameters were obtained from Ref. \cite{buckley2016structure}.}
\label{table:CaseSet}
\begin{tabular}{l l l l l l l l l l}
\hline
\textbf{} & \textbf{$C_{p}/u_{*}$}&\textbf{$C_{p}/U_{10}$}&\textbf{$u_{*}$}&\textbf{$U_{10}$}&\textbf{$C_{p}$}&\textbf{$a$}&\textbf{$\lambda_{p}$}&\textbf{$ak_{p}$}&\textbf{$f_{p}$}\\
 & & &(cm/s)&(m/s)&(m/s)&(cm)&(m)&{}&{(Hz)}\\
\hline
Wind waves & 3.7 & 0.12 & 16.7 & 5.00 & 0.62 & 0.50 & 0.25 & 0.13 & 2.5\\
Wind over swell & 31.7 & 0.97 & 4.1 & 1.34 & 1.30 & 2.00 & 1.08 & 0.12 & 1.2\\
\hline
\end{tabular}
\end{table}

\begin{figure}[h!]
\centering\includegraphics[width=0.7\linewidth]{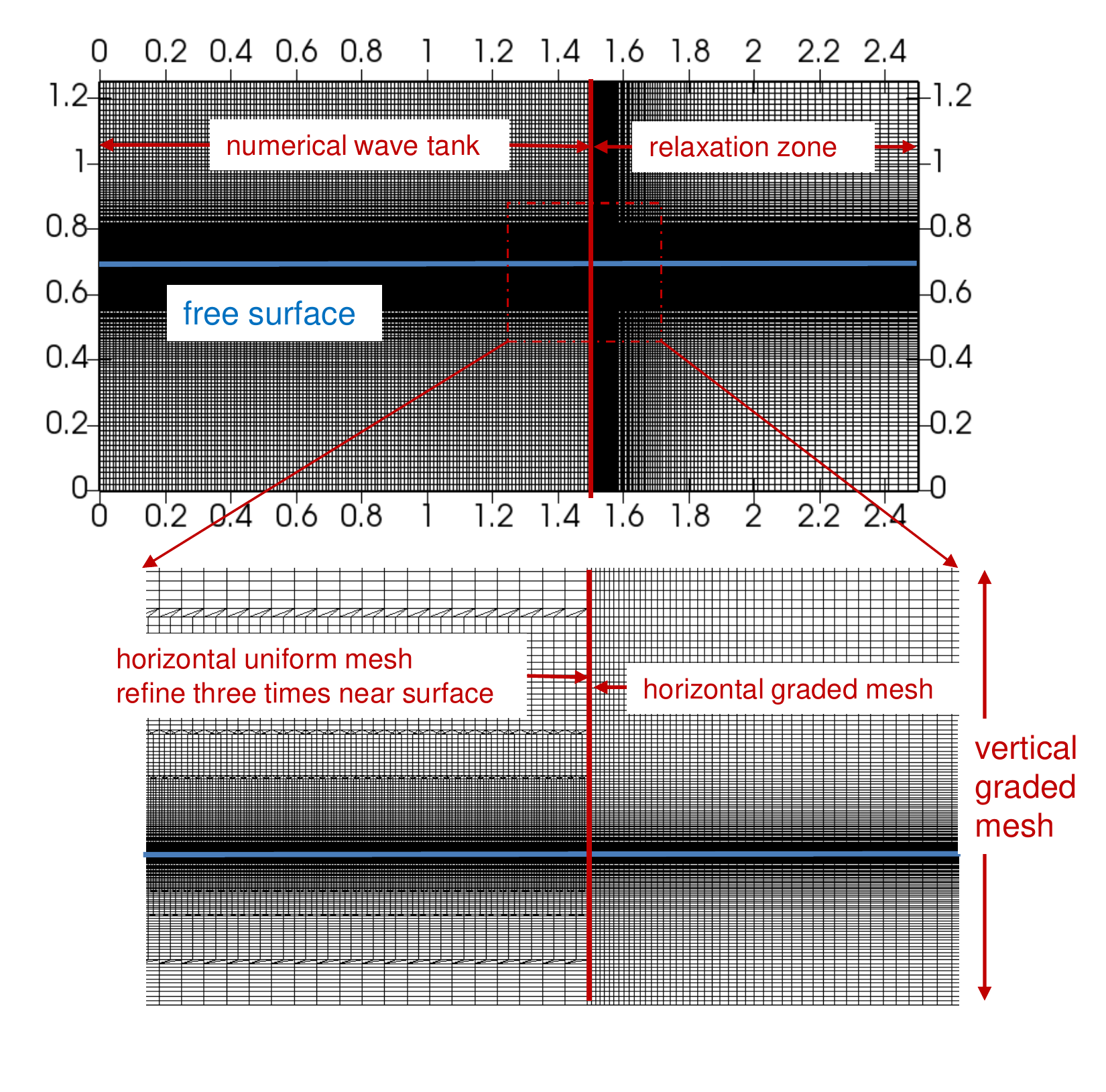}
\caption{Schematic diagram of mesh grid in the case of $C_{p}/u_{*}=3.7$}
\label{fig:mesh}
\end{figure}

In the 3D numerical wave tank, the airflow structure is measured with high resolution and compared with experimental data. To reduce the computational cost, the length of the numerical wave tank is set as around six times the wavelength. A relaxation region with a length of around four wavelengths is added at the end of the numerical wave tank to dissipate wave energy. The height (1.25 m) is the same as that of the experimental wave tank. The width is set as 0.1 m to reduce the computational cost. To accurately simulate wind field near the surface, finer meshes are used near the air-water interface, as shown in Fig. \ref{fig:mesh}. In the horizontal direction, the numerical fluid domain was meshed with a uniform grid. Grid refinement is applied in the vicinity of the free surface through refining the horizontal base grid three times. In the case of $C_{p}/u_{*}=31.7$, the total length of the wave tank is 10.4 m and the relaxation region is in the region of $6.4 <x<10.4$ m. The vertical resolution is 0.04 cm near the free surface. The base grid size in horizontal direction is 1 cm and the refined grid size is 0.125 cm. In the case of $C_{p}/u_{*}=3.7$, the total length of the wave tank is 2.5 m and the relaxation zone is in the region of $1.5 <x <2.5$ m. The vertical and horizontal resolutions are 0.02 cm and 0.125 cm near the free surface. 

In the experiment, the data was collected at a fetch of 22.7 m, where the wind waves were generated and mean wind profile follows the typical wall-bounded log-law profile. To be consistent with the experiment conditions, a regular wave is generated at the numerical wave tank inlet boundary with wave parameters defined in Table \ref{table:CaseSet}. Wind velocities at the inlet are prescribed via the summation of mean velocities and turbulent fluctuations. The mean velocities are defined following the law of wall profile using the parameters listed in Table \ref{table:CaseSet}. The turbulent fluctuations are produced using turbulent spot method \cite{kornev2007synthesis}. The turbulent variance $\overline{u'u'}$ is set as $4u_*^2$. The turbulent variance $\overline{v'v'}$ and $\overline{w'w'}$ are approximated using the ratios, $\sigma_v/\sigma_u$ and $\sigma_w/\sigma_u$, suggested by Counihan \cite{counihan1975adiabatic}.

\begin{equation}
\label{eq:18}
\sigma_u:\sigma_v:\sigma_w=1:0.75:0.5
\end{equation}

In the case of $C_p/u_*=3.7$, the Reynolds stresses ($ \overline{u'u'}$, $\overline{v'v'}, \overline{w'w'}$) are prescribed as (0.10, 0.06, 0.03) m$^2$/s$^2$. In the case of $C_p/u_*=31.7$, the Reynolds stresses are prescribed as (0.006, 0.003, 0.002) m$^2$/s$^2$. The integral length scale $^xL_u$ is set as 0.1 m, which is estimated as 0.2 of the height of the air passage gap. The integral length scales ($^yL_v$, $^zL_w$) are approximated as follows \cite{amerio2014numerical}.

\begin{eqnarray}
\label{eq:20}
^yL_v &=&0.3^xL_u\nonumber\\
^zL_w &=&0.2^xL_u
\end{eqnarray}

\begin{figure}[h!]
\centering\includegraphics[width=0.9\linewidth]{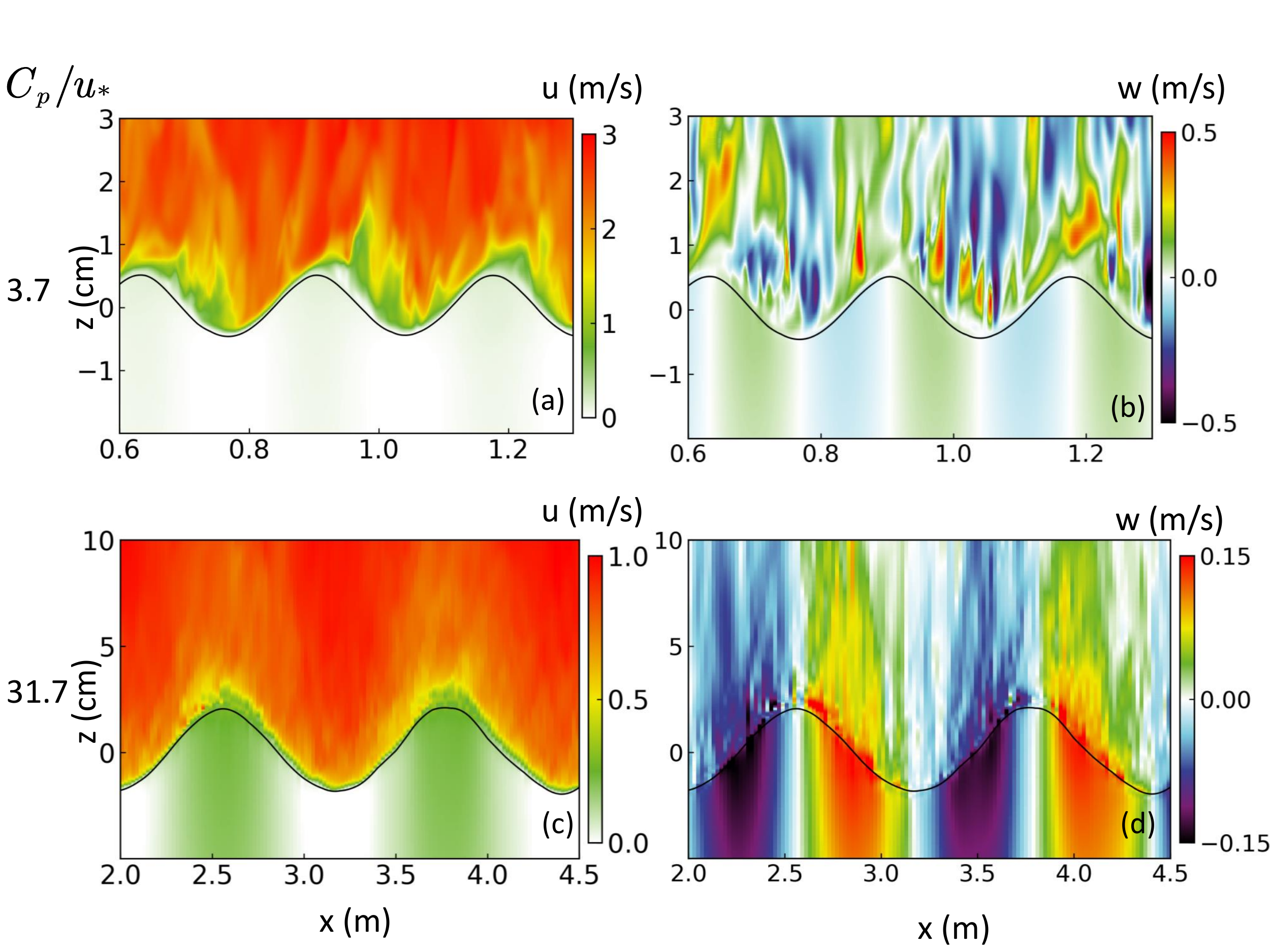}
\caption{(a) Instantaneous horizontal velocity fields $u$ in the case of $C_p/u_*=3.7$; (b) Instantaneous vertical velocity fields $w$ in the case of $C_p/u_*=3.7$; (c) Instantaneous horizontal velocity fields $u$ in the case of $C_p/u_*=31.7$; (d) Instantaneous vertical velocity fields $w$ in the case of $C_p/u_*=31.7$}
\label{fig:VelocityContour}
\end{figure}

For the 3D numerical wave tank, the front and back boundaries are set as a symmetrical condition. The top and bottom boundaries are treated as no-slip walls. With the maximum CFL number set as 0.5, the time step is $5e^{-5}$ s in the case of $C_p/u_*=3.7$ and $3e^{-4}$ s in the case of $C_p/u_*=31.7$. The wind wave field is simulated for 15 s in parallel with the entire domain divided into 384 subdomains. The contour plots of velocities, $u$ and $w$, at the last step for the two cases are shown in Fig. \ref{fig:VelocityContour}, where there are high intensity turbulences above the wave surface. In the case of $C_p/u_*=3.7$, airflow separation appears in the region past the crests of the waves, as shown in Fig. \ref{fig:VelocityContour}(a).

\begin{figure}[h!]
\centering\includegraphics[width=0.6\linewidth]{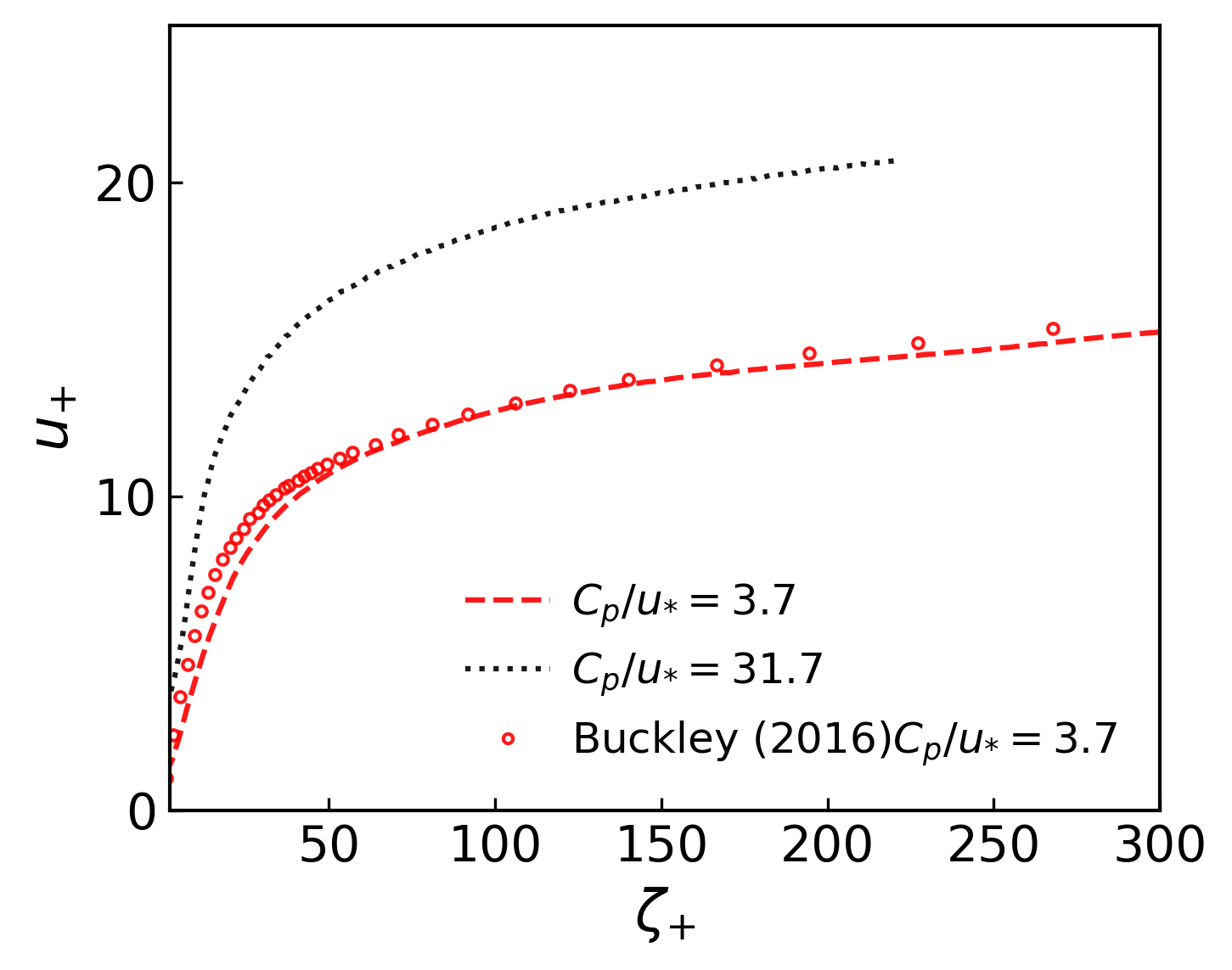}
\caption{Profiles of total averaged velocity $\bar u$ across all phases in surface-following coordinates.}
\label{fig:MeanVelocityProfile}
\end{figure}

Through conducting triple decomposition of the wind-wave velocity described in Section \ref{S:3}, the velocity field is decomposed into a mean velocity ($\bar{q}(\zeta)$) profile, a wave-coherent velocity field ($\tilde{q}(\xi,\zeta)$), and a turbulent velocity field ($q'(x,y,z,t)$). The decomposed numerical results are compared with measured results presented in \cite{buckley2016structure}. The profiles of the mean velocity $\bar u(\zeta)$ in surface following coordinates defined in Section \ref{S:3_2} are plotted in Fig. \ref{fig:MeanVelocityProfile}, where the quantities $\bar u(\zeta)$ and $\zeta$ are normalized as $u_{+}=\bar{u}/u_{*}$, $\zeta_{+}=\zeta u_{*}/\mu$. Fig. \ref{fig:MeanVelocityProfile} indicates that the simulated profile of the mean velocity $\bar u(\zeta)$ agrees well with the experimental data. In the region very close to the water surface ($\zeta_+ <50$), the simulated mean velocity is slightly smaller than the experimental results because the momentum in the air is transferred to the water and finer meshes are required to capture the very steep velocity gradient close to the air-water surface.

\begin{figure}[h!]
\centering\includegraphics[width=0.95\linewidth]{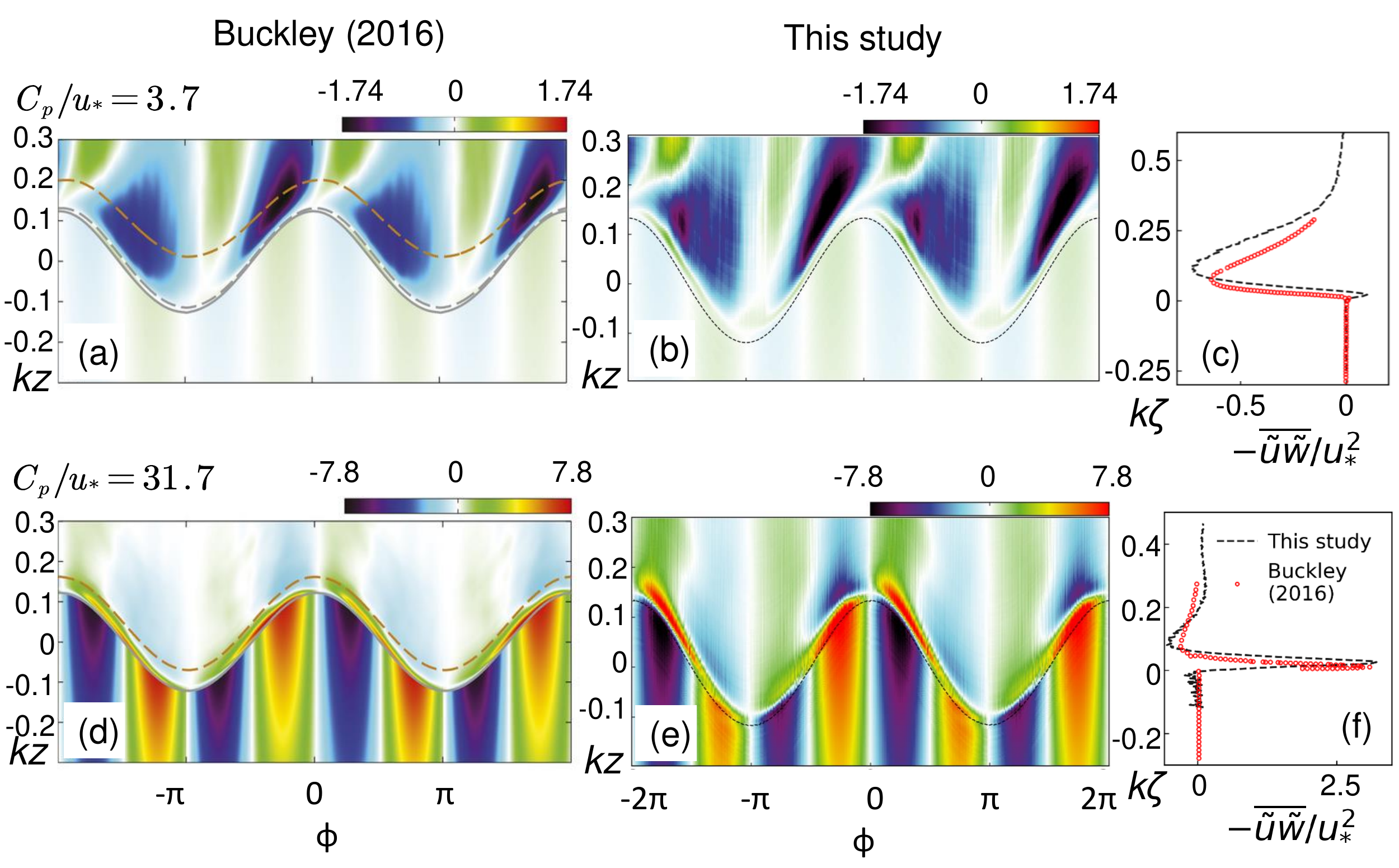}
\caption{Normalized phase averaged wave coherent stresses $-\langle\tilde{u}\tilde{w}\rangle /u_*^2$ from the experimental results in Ref. \cite{buckley2016structure}: (a) $C_p/u_*=3.7$, (d) $C_p/u_*=31.7$; Normalized phase averaged wave coherent stresses $-\langle\tilde{u}\tilde{w}\rangle /u_*^2$ in this study: (b) $C_p/u_*=3.7$, (e) $C_p/u_*=31.7$; Total mean wave coherent stress across all phases $-\overline{\tilde{u}\tilde{w}}/u_*^2$: (c) $C_p/u_*=3.7$, (f) $C_p/u_*=31.7$.}
\label{fig:waveInducedUW}
\end{figure}

In addition to the mean velocity, the simulated wave-induced momentum fluxes ($\tilde{u}\tilde{w}$) are compared with experimental results. The phase-averaged wave coherent stress ($-\langle\tilde{u}\tilde{w}\rangle /u_*^2$) and total mean wave stress ($-\overline{\tilde{u}\tilde{w}}/u_*^2$) are shown in Fig. \ref{fig:waveInducedUW}, where the phase-averaged velocity along two wavelengths is plotted to visualize the turbulence structure. Comparing Fig. \ref{fig:waveInducedUW}(a) and Fig. \ref{fig:waveInducedUW}(b), one can find that the distribution of phase-averaged wave coherent stress in the case of young wave ($C_p/u_*=3.7$) agrees well with experimental results in Ref. \cite{buckley2016structure}. The simulated distribution of phase-averaged wave coherent stress in the case of old wave ($C_p/u_*=31.7$) also agrees well with experimental results as shown in Fig. \ref{fig:waveInducedUW}(d) and Fig. \ref{fig:waveInducedUW}(e). Through comparing Fig. \ref{fig:waveInducedUW}(b) and Fig. \ref{fig:waveInducedUW}(e), one can find that the wave age has a significant effect on the wave coherent stresses. In the case of young wave, the wave coherent stresses are negative and asymmetric near the surface. In the case of old wave, the wave coherent stresses are positive and intense in a thin region close to the wave surface. The simulated total mean wave coherent stresses $-\overline{\tilde{u}\tilde{w}}/u_*^2$ across all phases are compared with experimental results in Fig. \ref{fig:waveInducedUW}(c) and Fig. \ref{fig:waveInducedUW}(f), where agreement between simulated and measured results can be observed in the case of young and old waves.

\begin{figure}[!htbp]
\centering\includegraphics[width=0.95\linewidth]{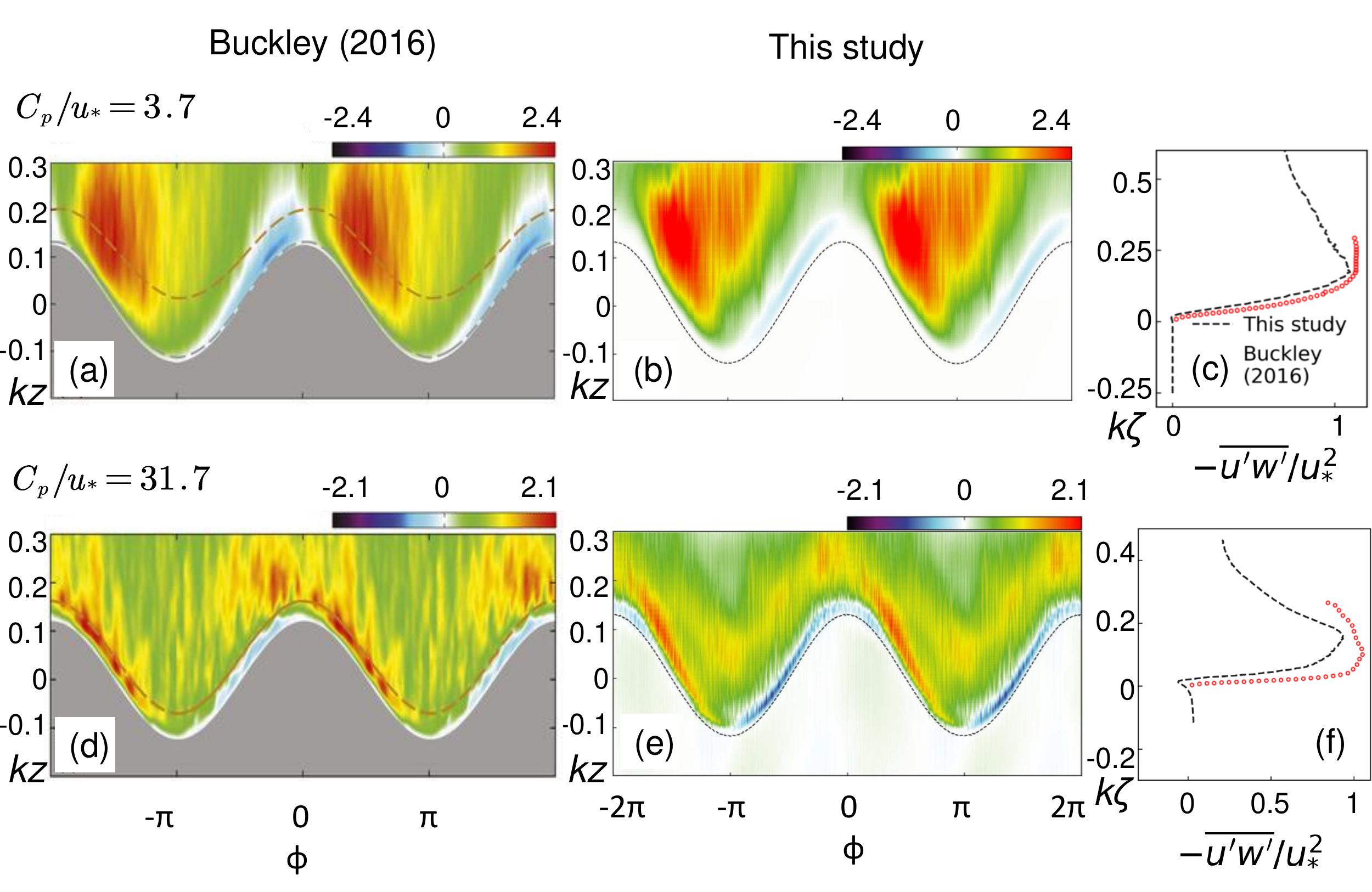}
\caption{Normalized phase average turbulent stress $-\langle u'w'\rangle /u_*^2$ from the experimental results in Ref. \cite{buckley2016structure}: (a) $C_p/u_*=3.7$, (d) $C_p/u_*=31.7$; Normalized phase average turbulent stress $-\langle u'w'\rangle /u_*^2$ in this study: (b) $C_p/u_*=3.7$, (e) $C_p/u_*=31.7$; Total mean turbulent stress across all phases $-\overline{u'w'}/u_*^2$: (c) $C_p/u_*=3.7$, (f) $C_p/u_*=31.7$.}
\label{fig:turbulentuw}
\end{figure}

Fig. \ref{fig:turbulentuw} shows the phase-averaged normalized turbulent stresses $-\langle u'w'\rangle /u_*^2$ and total averaged turbulent stresses across all phases $-\overline{u'w'}/u_*^2$ for the two wind wave cases. Comparison between Fig. \ref{fig:turbulentuw}(a) and Fig. \ref{fig:turbulentuw}(b) and comparison between Fig. \ref{fig:turbulentuw}(d) and Fig. \ref{fig:turbulentuw}(e) show that the distributions of the turbulent stresses are consistent with that observed in the experiment. As shown in Fig. \ref{fig:turbulentuw}(b) and Fig. \ref{fig:turbulentuw}(e), the phase-averaged turbulent stresses vary significantly with wave ages. Over young waves, the turbulent stresses are strong and positive downwind of the waves, which is caused by the air separation that occurs downwind of waves. Over old waves, the positive turbulent stresses are not that strong as that over young waves. Intense turbulent stresses occur close to the wave surface. Also, over old waves, the positive stress region extends to the upwind face of the wave which is confined to the downwind side over young waves. The average turbulent stresses across all phases are shown in Figs. \ref{fig:turbulentuw}(c) and (f), where good agreement can be observed between the simulated result and experimental result from \cite{buckley2016structure}. 

\FloatBarrier

Sections \ref{S:4.1} and \ref{S:4.2} verify the two-phase numerical model via comparing simulated results of wave elevations under wind forcing and wind field structure above wave surface with the corresponding experimental data. The following section will apply the verified numerical model to characterize the coupled wind-wave field on a realistic scale.

\section{Numerical case study}

In this section, the numerical model is applied to analyze the characteristics of coupling wind-wave fields with high wind speed and wave height on a realistic scale. The computational domain has a size of 450 m $\times$ 160 m $\times$ 200 m with a water depth of 100 m. The relaxation zone is within $300<x<450$ m. Vertically graded meshes are generated with a minimal grid size of 0.06 m near the water surface and a maximum grid size of 5 m away from the water surface. In the horizontal direction, the numerical fluid domain was meshed with a uniform grid size of 4 m. The meshes near the free surface ($85<z<135$ m) are refined. The minimal mesh size in horizontal direction is 0.5 m. The total number of meshes of the computational domain is around 20 million. 

As mentioned before, the present study simulates two wind-wave scenarios: regular wave with uniform wind and regular wave with highly turbulent wind. At the inlet boundary, the water phase velocity is defined through the prescribed regular wave and the air phase velocity is defined as turbulent or uniform wind speed. The parameters for the prescribed uniform wind and waves are listed in Table \ref{table:CaseStudy}, where the value of $U_{10}$ is chosen to represent wind speeds in typical tropical cyclones. Based on the wave parameters, stokes nonlinear waves are generated at the inlet. The case 1, case 2 and case 3 correspond to the situations of different wave ages and represent the scenarios of gentle wind forcing and extreme hurricane wind forcing. In the case 4, the wind condition is same as that in case 1 but the wave height is lower than that in case 1. The parameters for generating the wind turbulences, such as the integral length scales, turbulence intensities, and Reynolds stress, are determined referring to the ESDU 85020 \cite{esdu2001characteristics}. Turbulent wind velocities are generated at the inlet boundary using the corresponding wind field parameters listed in Table \ref{table:CaseStudy}. The boundary conditions are defined the same as that for the 3D numerical wave tank in section \ref{S:4.2} except the conditions at the top boundary, where a fixed shear stress ($\tau = \rho u_*^2$) is applied to maintain the inlet profiles over a long distance. Simulations for all the cases are carried out for 120 s in parallel with the entire domain divided into 512 subdomains and the maximum CFL number set as 0.5. Statistics near the air-water surface are obtained after 60 s with a time interval of 0.01 s.

\begin{table}[h]
\centering
\caption{Parameters for wind and wave generated at the inlet boundary}
\label{table:CaseStudy}

\begin{tabular}{l l l| l l l |l l l l l l}
\hline
\multicolumn{3}{c|}{Case} & \multicolumn{3}{c|}{Logarithmic wind conditions} & \multicolumn{4}{c}{Wave conditions}\\
\hline
{No.} &\textbf{$C_{p}/u_{*}$} & \textbf{$C_{p}/U_{10}$} & \textbf{$U_{10}$}&\textbf{$z_0$}&\textbf{$u_{*}$}&\textbf{$C_p$}&\textbf{$a$}&\textbf{$\lambda$}&\textbf{$f$}\\
& & &(m/s) & (cm) &(m/s)&(m/s)&(m)&(m)&{(Hz)}\\
\hline
1 & 6.95 & 0.264 & 30 & 0.02 & 1.137 & 7.9 & 2.00 & 40 & 0.2\\
2 & 20.84 & 0.791 & 10 & 0.02 & 0.379 & 7.9 & 2.00 & 40 & 0.2\\
3 & 4.17 & 0.158 & 50 & 0.02 & 1.895 & 7.9 & 2.00 & 40  & 0.2\\
4 & 6.95 & 0.264 & 30 & 0.02 & 1.137 & 7.9 & 1.00 & 40 & 0.2\\
\hline
\end{tabular}
\end{table}

\subsection{Wind effect on wave elevation}
\begin{figure}[h!]
\centering\includegraphics[width=0.9\linewidth]{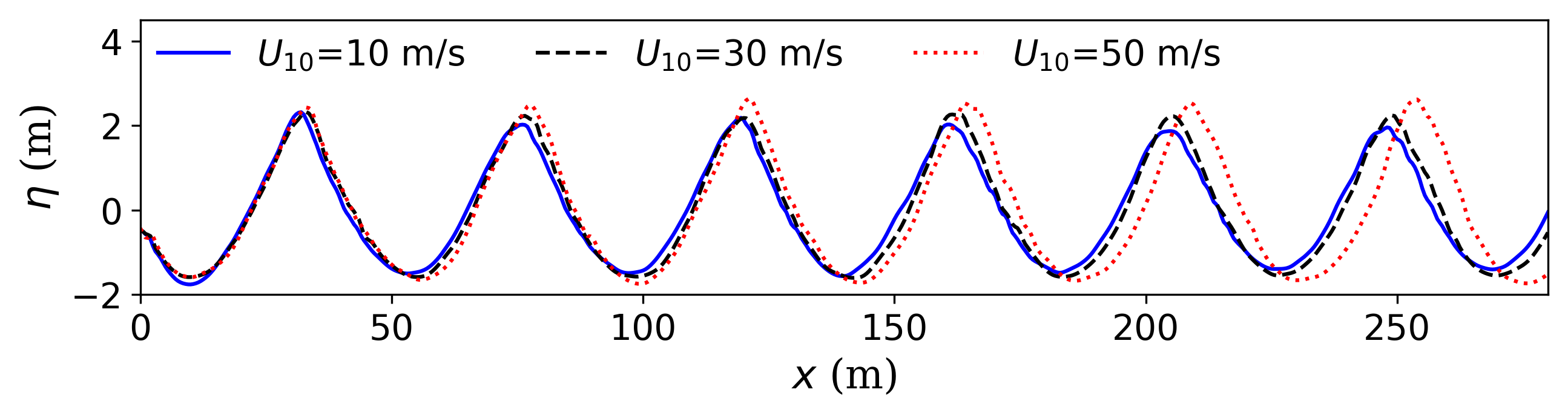}
\caption{Evolution of wave elevations under different following wind speeds}
\label{fig:WaveElevationCompare}
\end{figure}

As shown in Table \ref{table:CaseStudy}, three following wind speeds $U_{10}=$ 10, 30, and 50 m/s with an identical wave profile are simulated. Fig. \ref{fig:WaveElevationCompare} shows the spatial distribution of wave profiles at $t=120$ s. As shown in the Fig. \ref{fig:WaveElevationCompare}, the wave profiles are asymmetry, with a larger crest than the trough, which indicates the typically nonlinear character. One can observe that as the following wind speed increases, the wave profile shifts forward and the wave height increases. 

\subsection{Wind field evolution along wave direction}

\begin{figure}[h!]

\centering\includegraphics[width=0.8\linewidth]{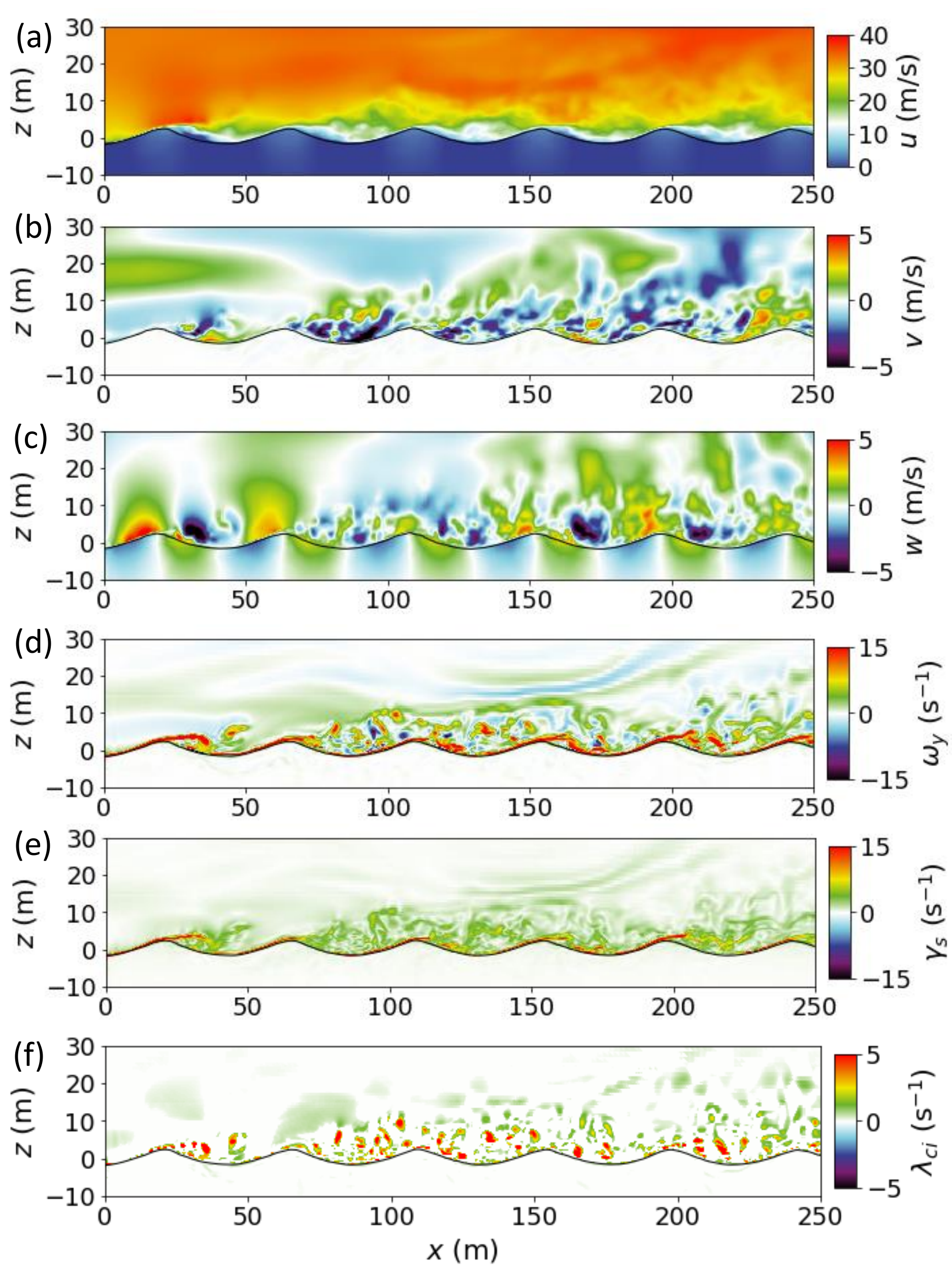}
\caption{Instantaneous fields in case 1 with turbulences added at the inlet boundary. (a) Streamwise (in x direction) velocity $u$ field; (b) Spanwise (in y direction) velocity $v$ field; (c) Vertical velocity $w$ field; (d) Vorticity field in spanwise direction, $\omega_y$; (e) Shearing strength $\gamma_s$ field; (f) Swirling strength $\lambda_{ci}$ field.} 
\label{fig:Velocity30}
\end{figure}

To capture the wind field evolution as wave propagates, the instantaneous velocity field in XZ plane at the last step in case 1 is shown in Fig. \ref{fig:Velocity30}. Fig. \ref{fig:Velocity30}(a) illustrates clear air separation past the wave crest. The streamwise velocities decreases downwind of the wave crest. Fig. \ref{fig:Velocity30}(b) shows the instantaneous velocities in spanwise direction, which can be regarded as turbulence because the spanwise velocity has a zero mean. Fig. \ref{fig:Velocity30}(c) shows the vertical velocity. From the velocity contours, it can be seen that along the wave, the wind turbulence increases and small-scale turbulences are generated near the wave surface. The vorticity field is shown in Fig. \ref{fig:Velocity30}(d), where a layer with high positive vorticity can be observed close to the upwind surface of the wave. This high vorticity corresponds to the shear effect between air and water, as shown in Fig. \ref{fig:Velocity30}(e), which shows the field of shearing strength $\gamma_s$. The shearing strength is introduced in Ref. \cite{zhang2008asymmetric} to measure the strength of anisotropic stretching, which is defined as,

\begin{equation}
\label{eq:21}
\gamma_s = \frac{1}{2}\sqrt{\left (\frac{\partial u}{\partial x}-\frac{\partial w}{\partial z}\right)^2
+\left(\frac{\partial u}{\partial z}+\frac{\partial w}{\partial x}\right)^2}
\end{equation}
The high positive shear-related vorticity become unstable after passing the crest of the wave and some small centers of vortity appear. The small centers of vortity correspond to the vortical vorticity, as shown in Fig. \ref{fig:Velocity30}(f), which illustrates the fields of swirling strength. The swirling strength is introduced in Ref. \cite{zhou1999mechanisms} to measure the strength of the local swirling motion, which is defined as the imaginary part of complex eigenvalues for the velocity gradient tensor,

\begin{equation}
\label{eq:23}
\lambda_{ci} = \operatorname{Im}\Bigg\{\frac{1}{2}\left[\sqrt{\left (\frac{\partial u}{\partial x}-\frac{\partial w}{\partial z}\right)^2
+\left(\frac{\partial u}{\partial z}+\frac{\partial w}{\partial x}\right)^2}\right]\Bigg\}
\end{equation}
The swirling strength is an effective vortex indicator. From Fig. \ref{fig:Velocity30}(f), it can be seen that vortexes, centered at the local extrema of high swirling strength, are produced after wind passes the wave crests. Therefore it can be concluded that turbulences are strengthened when the free shear layers detach from the wave surface after passing the wave crest.

\begin{figure}[h!]

\centering\includegraphics[width=0.9\linewidth]{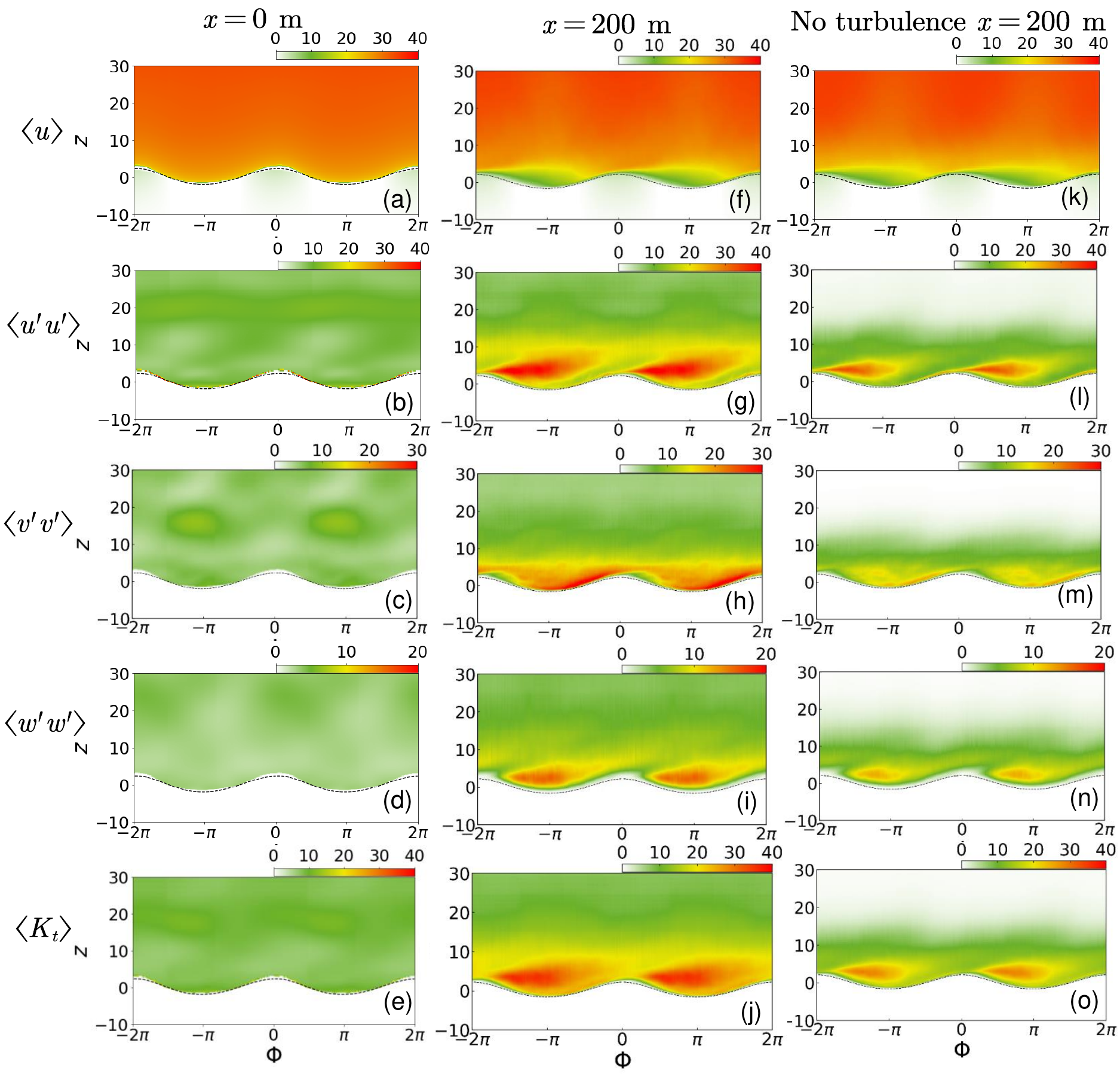}
\caption{Phase averaged velocity field in case 1 with $U_{10}$ = 30 m/s and H=4 m. (a,f,k) Phase averaged streamwise velocity $<u>$; (b,g,l) Phase averaged turbulent variance $<u'u'>$; (c,h,m) Phase averaged turbulent variance $<v'v'>$; (d,i,n) Phase average turbulent variance $<w'w'>$; (e,j,o) Phase averaged turbulence kinetic energy $<K_t>$.}
\label{fig:inletOutletCompare_U30H4_contour}
\end{figure}

To study the interaction between turbulent winds and surface waves, the phase-average method is used to quantify the statistical properties of wind turbulence. Using the phase average method presented in Section \ref{S:3_2}, the wind turbulence structures are illustrated along two wavelengths to clearly visualize the characteristic evolution. To further analyze the wind velocity evolution, the phase averaged streamwise velocity ($<u>$), turbulent variances ($<u'u'>$, $<v'v'>$, and $<w'w'>$), and turbulence kinetic energy ($<K_t>$) at $x=0$ m and $x=200$ m in case 1 are presented in the first and second column in Fig. \ref{fig:inletOutletCompare_U30H4_contour}. For comparison, the phase averaged velocity field in the case of uniform wind without adding turbulences are simulated and plotted in the third column of Fig. \ref{fig:inletOutletCompare_U30H4_contour}. Comparing the phase averaged velocities $u$ at $x=0$ m and $x=200$ m, one can find that the wave decreases the averaged streamwise velocity downwind of the waves, which is caused by the sheltering effect. It is consistent with that observed in the instantaneous velocity $u$ fields, as shown in Fig. \ref{fig:Velocity30} (a). Comparing the phase averaged turbulent variances ($<u'u'>$, $<v'v'>$, and $<w'w'>$) at $x=0$ m and $x=200$ m, one can find that the wave enhances wind turbulence near the wave surface. An intense turbulent variance $<u'u'>$ region can be observed past the crest of the waves. The regions with maximum variances, $<u'u'>$ and $<v'v'>$, happen at the downwind faces of waves while the regions with maximum variances $<w'w'>$ happen near the upwind side of the waves. Furthermore, one can find that the phase averaged turbulence kinetic energy $<K_t>$ has an intense turbulence region past the crest of the waves, which is similar to the turbulent variance $<u'u'>$ because $<K_t>$ is mainly contributed by the variances $<u'u'>$ in the stream-wise direction. In the case without turbulence, the phase averaged velocity $<u>$ is almost the same as that with turbulence added at the inlet boundary, and the phase averaged turbulent variances at $x=200$ m show similar features yet with smaller amplitude. 

\begin{figure}[h!]

\centering\includegraphics[width=0.9\linewidth]{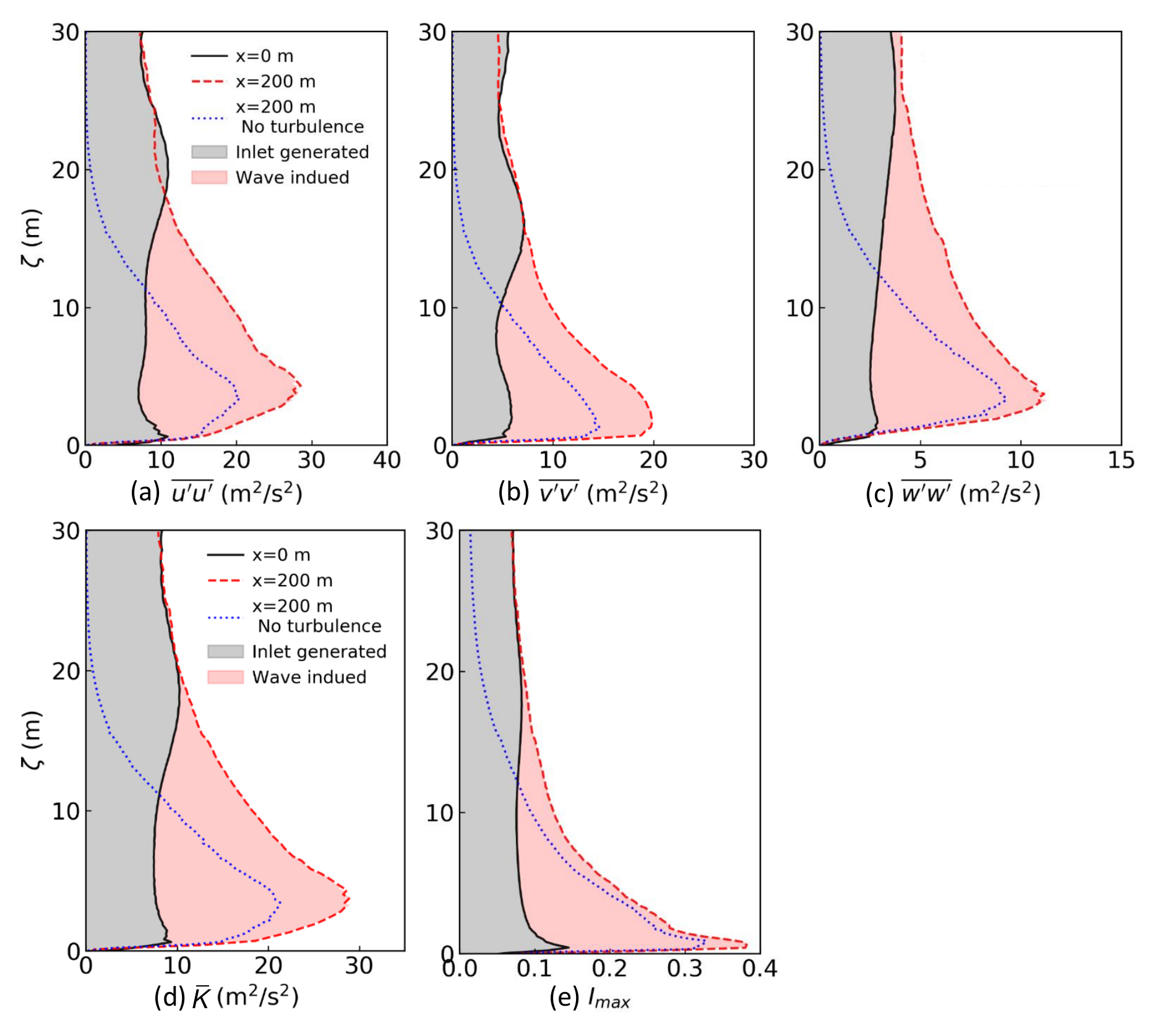}.
\caption{Total mean (across all phases) turbulent variances (a) $\overline{u'u'}$, (b) $\overline{v'v'}$, and (c) $\overline{w'w'}$. (d) Total mean turbulence kinetic energy $\overline{K}$. (e) Maximum turbulence intensity $I_{max}$ across all phases.}

\label{fig:inletOutletCompare_U30H4_verticalAverage}
\end{figure}

To quantify the generated turbulence and wave induced turbulence, the vertical averaged variances ($\overline{u'u'}$, $\overline{v'v'}$, and $\overline{w'w'}$) and vertical averaged turbulence kinetic energy ($\overline{K}$) for all phases are shown in Figs. \ref{fig:inletOutletCompare_U30H4_verticalAverage}(a), (b), (c) and (d). The vertical averaged quantities are averaged with respect to the surface following vertical coordinate $\zeta$. The maximum turbulent intensity, $I_{max}$, is plotted in Fig. \ref{fig:inletOutletCompare_U30H4_verticalAverage} (e), which is defined as,
\begin{equation}
\label{eq:23_2}
I_{max}=\frac{\sqrt{\frac{2}{3}<K_t>_{max}}}{\overline{U}}
\end{equation}
where $<K>_{max}$ is the maximum value across all phases. The black lines represent the vertical profiles of turbulent variances at $x=0$ m, which is the generated turbulence at the inlet boundary. The red dashed lines represent the turbulent variances at $x=200$ m with turbulence added at the inlet boundary. The blue dotted lines denote the turbulent variances at $x=200$ m in case without adding turbulence at the inlet boundary. Comparing the turbulence at $x=0$ m and at $x=200$ m indicates that the waves induce turbulence near the wave surface and increase the turbulent variances. The added turbulence at the inlet boundary is marked in grey and the wave induced turbulence is marked in red.  At 200 m from the inlet, the streamwise variance ($\overline{u'u'}$) peak increases to 29 m$^2$/s$^2$. The spanwise and vertical components increase to 20 m$^2$/s$^2$ and 11 m$^2$/s$^2$ respectively. At a height of 5 m, the total averaged turbulence kinetic energy $\overline{K}$ increases from 8 m$^2$/s$^2$ to 27 m$^2$/s$^2$, and the $I_{max}$ increases from 8$\%$ to 21$\%$. From the inlet to $x=200$ m, the wave strengthened the wind turbulence by more than 100$\%$. In Fig. \ref{fig:inletOutletCompare_U30H4_verticalAverage}, the difference between the case with turbulence added and without adding turbulence is mainly caused by the added turbulence. The turbulence at the target location is the summation of the added turbulence and wave induced turbulence. The waves influence the wind fields dramatically near the wave surface and increase the wind turbulence up to a height of 20 m  above the wave surfaces.

\begin{figure}[h!]

\centering\includegraphics[width=0.9\linewidth]{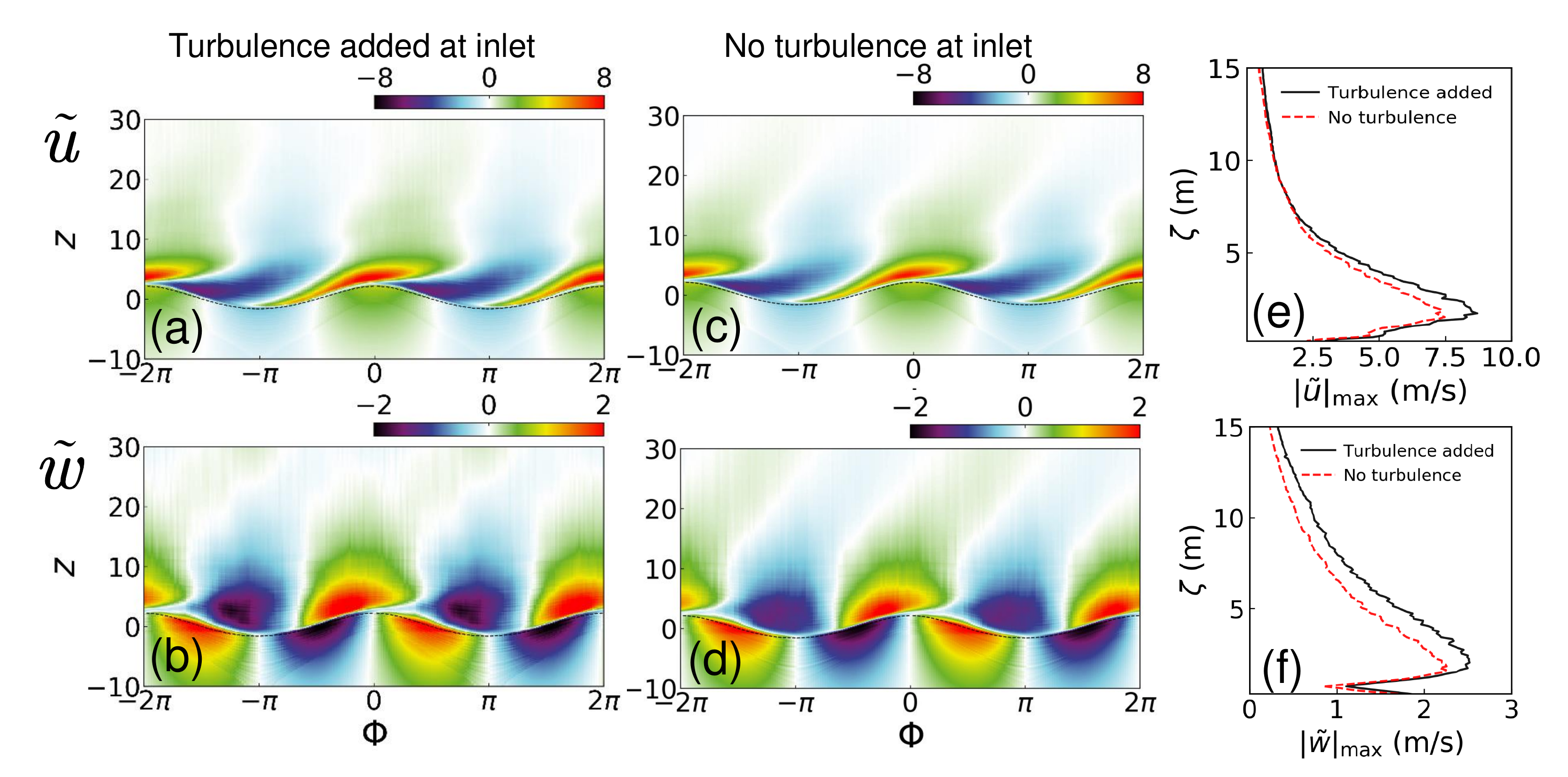}.
\caption{(a,c) Wave-coherent streamwise velocity $\tilde{u}$. (b,d) Wave-coherent vertical velocity $\tilde{w}$. (e) Vertical profile of maximum wave-coherent streamwise velocity across all phases $|\tilde{u}|_{max}$. (f) Vertical profile of maximum wave-coherent vertical velocity across all phases $|\tilde{w}|_{max}$.}

\label{fig:waveCoherent_U30H4}
\end{figure}

In addition to its effect on wind turbulence, the wave surface profiles influence the averaged wind field as shown in Fig. \ref{fig:inletOutletCompare_U30H4_contour}(f). The variance of phase averaged wind velocity with wave phase can be represented as wave coherent velocity ($\tilde{u}$), which is obtained as $\tilde{u}=<u>-\overline{u}$. The wave coherent velocities $\tilde{u}$ and $\tilde{w}$ at $x=200$ m are plotted in Fig. \ref{fig:waveCoherent_U30H4}(a) and Fig. \ref{fig:waveCoherent_U30H4}(c). Near the wave surface, the wave coherent velocity $\tilde{u}$ is positive along upwind surface and negative along the downwind surface. As wave height increases, the $\tilde{u}$ contours 
tilts downwind. The wave coherent velocity $\tilde{w}$ is positive at the upwind face and negative at the downwind face. Comparing to the $\tilde{w}$ contour under wave surface, the $\tilde{w}$ contour above wave surface exhibits an opposite negative-positive pattern. Due to the sheltering effect, the negative-positive pattern of $\tilde{w}$ above wave surface is phase-shifted upwind. The wave coherent velocities are dominant near the wave surface and negligible away from the surface. The wave coherent velocities $\tilde{u}$, $\tilde{w}$ in the case without turbulence added at the inlet boundary are plotted in Fig. \ref{fig:waveCoherent_U30H4}(c) and \ref{fig:waveCoherent_U30H4}(d). Through comparing contour plots of $\tilde{u}$ between Fig. \ref{fig:waveCoherent_U30H4}(a) and Fig. \ref{fig:waveCoherent_U30H4}(c), one can find that the patterns of $\tilde{u}$ show common features with or without adding turbulence. The phenomenon is also observed in the contour plots of $\tilde{w}$. The maximum absolute wave coherent velocities across all phases ($|\tilde{u}|_{max}$, $|\tilde{w}|_{max}$) are shown in Fig. \ref{fig:waveCoherent_U30H4}(e) and Fig. \ref{fig:waveCoherent_U30H4}(f). The black lines denote the results in the case with adding turbulence. The red dashed lines denote the results in the case without adding turbulence. In case 1 with $a=2$ m and $U_{10}=30$ m/s, the wave induces wave coherent velocities to 8.7 m/s in the streamwise direction and 2.5 m/s in the vertical direction. The maximum wave coherent velocities $|\tilde{u}|_{max}$ and $|\tilde{w}|_{max}$ are smaller than 1.5 m/s ($0.05U_{10}$) above heights of 8 m and 6 m, respectively, which indicates that the wave profiles influence the averaged wind fields $\tilde{u}$ and $\tilde{w}$ to heights of 8 m and 6 m. Through comparing the black and red lines in Fig. \ref{fig:waveCoherent_U30H4}(e) and Fig. \ref{fig:waveCoherent_U30H4}(f), one can find that the wave coherent velocity $\tilde{u}$ and $\tilde{w}$ in the case with adding turbulence are slightly larger than that without adding turbulence. Comparing Fig. \ref{fig:waveCoherent_U30H4}(b) and Fig. \ref{fig:waveCoherent_U30H4}(d) shows more intense of negative $\tilde{w}$ in case with turbulence added, which indicates that the added turbulence enhances the sheltering effect.

\subsection{Influence of wind velocities and wave heights on wind structures}
The previous section indicates that wave surfaces influence the wind fields through inducing turbulences and changing the structure of averaged wind fields. In this section, the wind field under different wind velocities ($U_{10}$=10, 30, and 50 m/s) and wave height ($H$ = 4, and 2 m) are compared. The wind and wave conditions are listed in Table. \ref{table:CaseStudy}.   

\begin{figure}[htbp!]

\centering\includegraphics[width=0.9\linewidth]{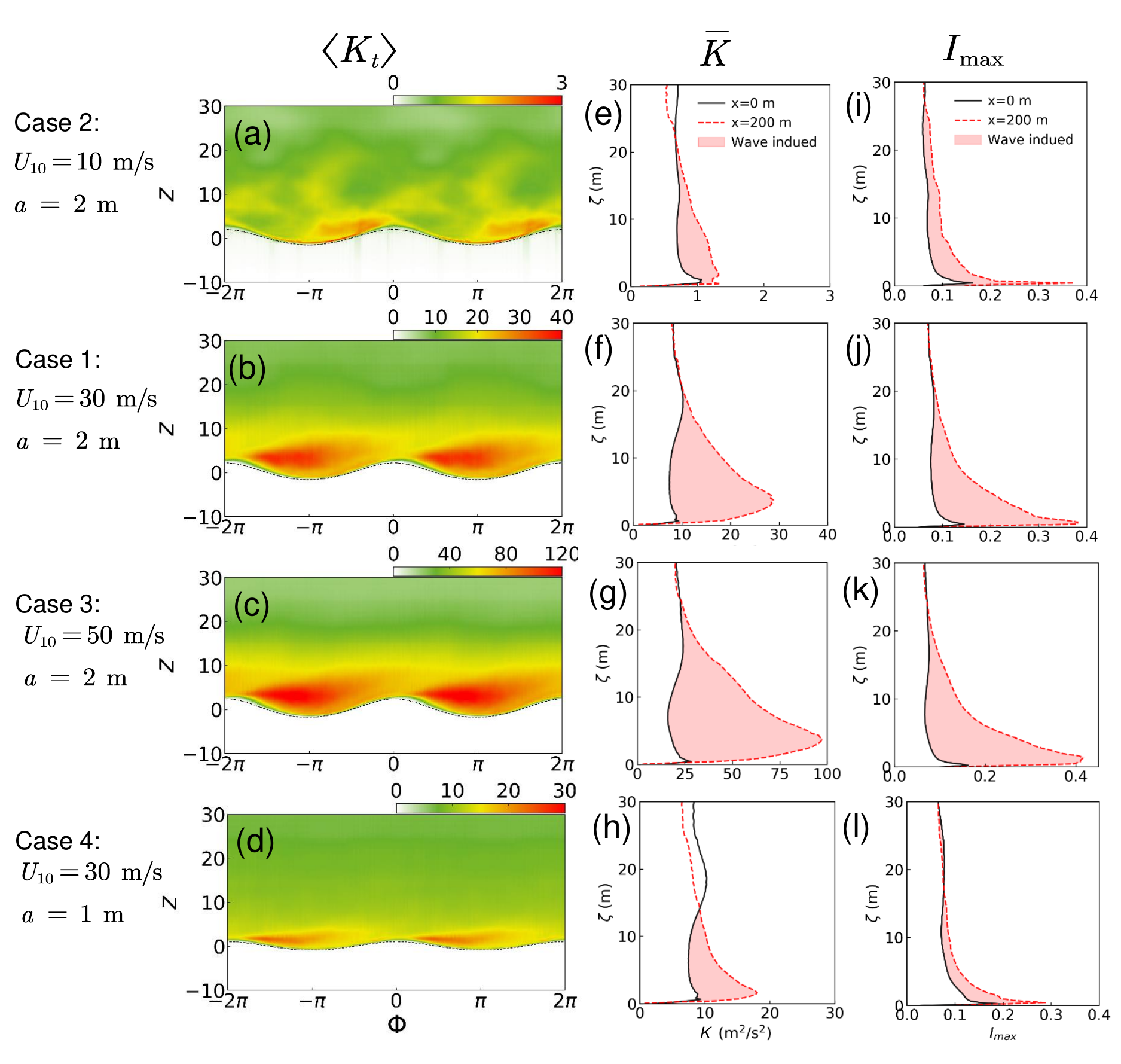}.
\caption{(a,b,c,d)Phase averaged turbulence kinetic energy $<K_t>$; (e,f,g,h) Total mean turbulence kinetic energy $\overline{K}$; (i,j,k,l) Maximum turbulence intensity $I_{max}$ across all phases.}

\label{fig:caseStudy_turbulence}
\end{figure}

\begin{figure}[htbp!]

\centering\includegraphics[width=0.4\linewidth]{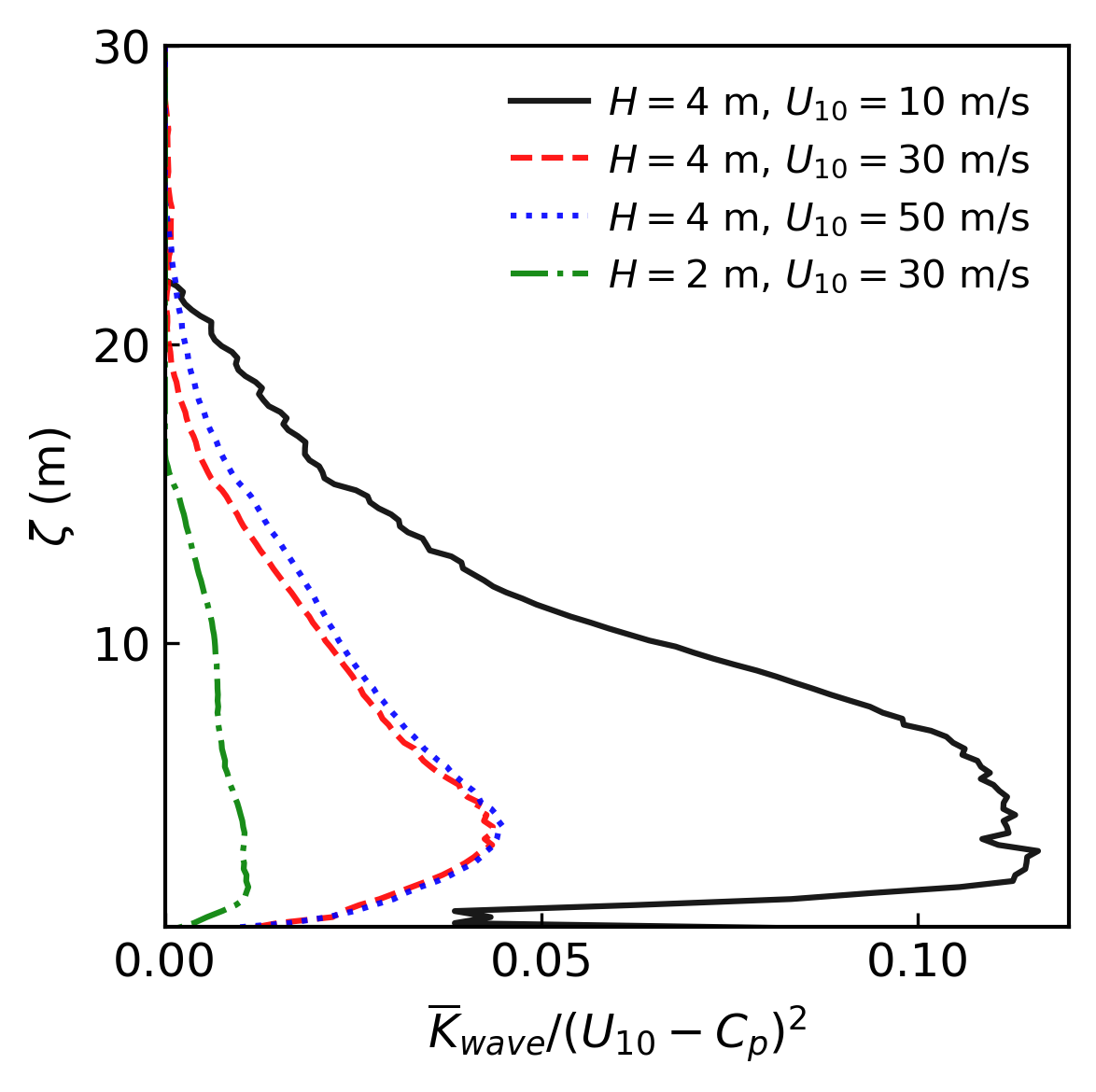}.
\caption{Normalized total mean wave induced turbulence kinetic energy $\overline{K}_{wave}/(U_{10}-C_p)^2$.}

\label{fig:wave_induced_turbulence}
\end{figure}

Fig. \ref{fig:caseStudy_turbulence} shows the wind turbulence obtained in the four cases. The first column shows the phase averaged turbulence kinetic energy ($<K_t>$) at $x$ = 200 m. The second and third columns illustrate the vertical profile of total mean turbulence kinetic energy ($\overline{K}$) and the maximum turbulent intensity ($I_{max}$) across all phases at the inlet boundary, $x=0$ m, and $x=200$ m, respectively. In case 2 with $U_{10}$ = 10 m/s and $H$ = 4 m, the regions with intense turbulent kinetic energy are located at the upwind side of the waves, as shown in the Fig. \ref{fig:caseStudy_turbulence}(a). While in case 1, the intense turbulent $<K_t>$ regions locate at the downwind side of the waves. The different pattern in Fig. \ref{fig:caseStudy_turbulence}(a) is caused by a "reversed sheltering effect" \cite{buckley2016structure}. In case 2, with the wave age $C_{p}/u_{*}$ = 20.84, the wind blows over fast moving old waves. Near the wave surfaces, the waves travel faster than the winds and the sheltering effect takes place at the upwind faces of the waves. As the wind velocity $U_{10}$ is close to the wave phase velocity, the wave-generated wind turbulence is smaller than that in case 1. At a height of 5 m, the vertical averaged turbulence kinetic energy $\overline{K}$ increases from 0.8 m$^2$/s$^2$ to 1.2 m$^2$/s$^2$, and $I_{max}$ increases from 8$\%$ to 11$\%$. Under high wind forcing ($U_{10}=50$ m/s), the pattern of $<K_t>$ is the same as that in case 1. In case 3, the wind velocity is much higher than $C_p$ and intense turbulences are generated above the wave surface. At a height of 5 m, the vertical averaged $\overline{K}$ increases from 17 m$^2$/s$^2$ at the inlet boundary to 89 m$^2$/s$^2$ at $x=200$ m, and $I_{max}$ increases from 7$\%$ to 24$\%$. Comparison between the second and third rows in Fig. \ref{fig:caseStudy_turbulence} shows that identical wave height ($H$=4 m) has close influential zone (roughly 20 m) in the wind field, and higher wind forcing speed causes stronger turbulences. In case 4, similar patterns of intensified wind turbulence past the waves are shown in Fig. \ref{fig:caseStudy_turbulence}(d). Through comparison between case 4 and cases 1 through 3, one can find that lower wave heights induce weaker turbulences and affect the wind field to a lower region. At the height of 5 m, the vertical averaged turbulence kinetic energy $\overline{K}$ increases from 7 m$^2$/s$^2$ to 13 m$^2$/s$^2$, and $I_{max}$ increases from 8$\%$ to 11$\%$. As shown in Figs. \ref{fig:caseStudy_turbulence} (h) and (l), the lower wave height in case 4 intensifies the wind turbulence to a height of around 15 m. Fig. \ref{fig:wave_induced_turbulence} shows the normalized total mean wave-generated wind turbulence kinetic energy $\overline{K}_{wave}/(U_{10}-C_p)^2$. One can find that the normalized wave-generated wind turbulence kinetic energy is the same in case 1 and case 3, which indicates that the wave-generated wind turbulence is proportional to the relative speed between wind velocity and wave phase speed under young wave conditions. The black line is different from the red and blue lines because the wave travels faster than the wind in case 2. Comparison between the red and green lines shows that the normalized  wave-generated turbulence is smaller above lower height waves.

\FloatBarrier

\FloatBarrier

\begin{figure}[!htb]

\centering\includegraphics[width=0.9\linewidth]{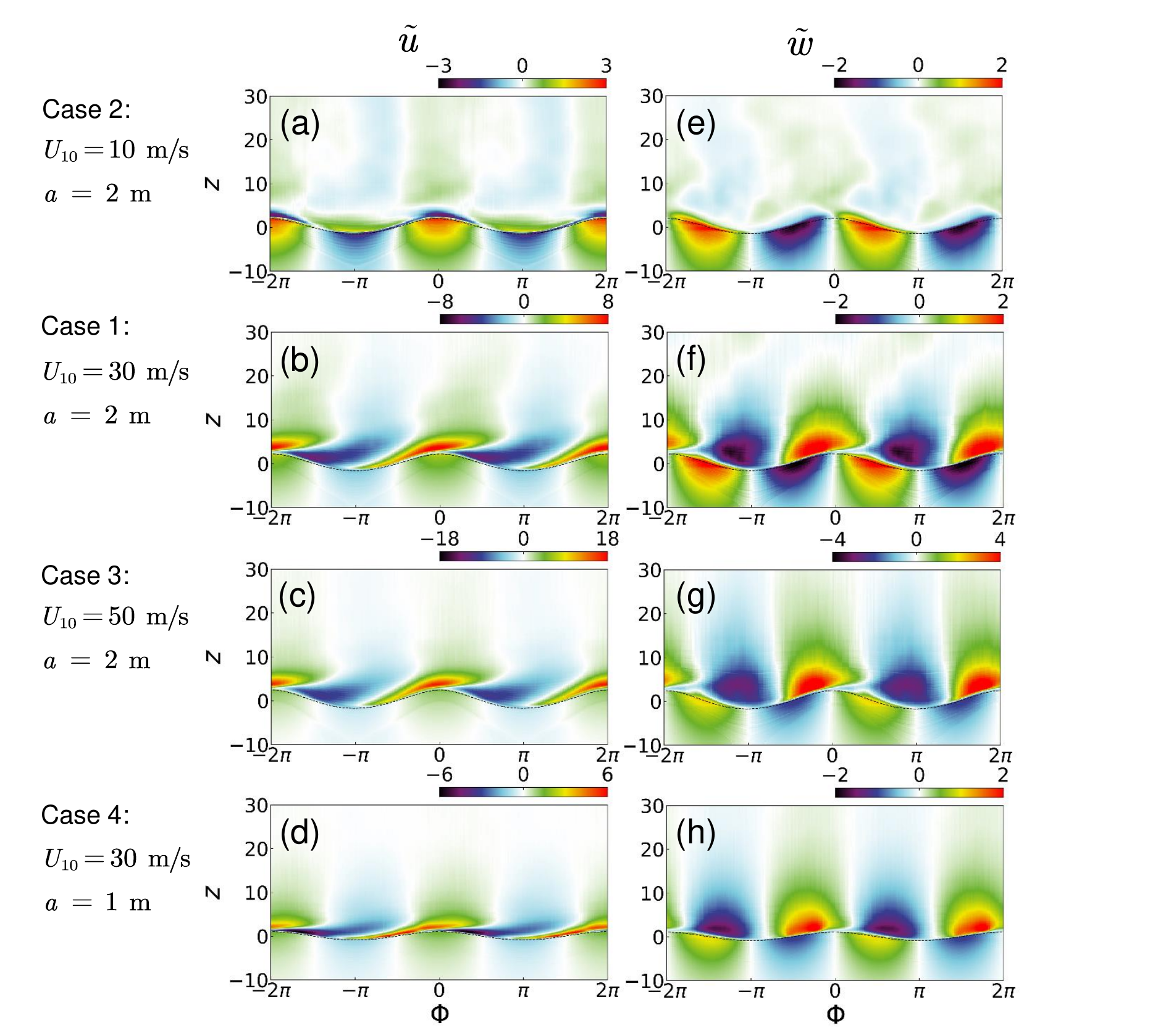}.
\caption{Wave induced velocity field in the four cases: (a-d) Wave coherent velocity in streamwise direction $\tilde{u}$; (e-h) Wave coherent velocity in vertical direction $\tilde{w}$.}

\label{fig:caseStudy_waveCoherence}
\end{figure}

Fig. \ref{fig:caseStudy_waveCoherence} shows the wave coherent velocities $\tilde{u}$ and $\tilde{w}$ in cases 1 through 4. In case 2 with $U_{10}=10$ m/s and $a=2$ m, the wave coherent velocity $\tilde{u}$ is positive above troughs and negative above crests near the wave surface, which is opposite to that in case 1. This opposite feature is due to the lower wind velocity near the wave surface relative to the wave phase speed. The wave coherent velocity $\tilde{w}$ close to the wave surface matches well with the water particle velocities, with negative $\tilde{w}$ at the upwind side of waves and positive at the downwind side. The intensified regions of wave coherent velocities are very close to the wave surfaces. In the case of wind travelling over old waves, the wave coherent velocities $\tilde{u}$ and $\tilde{w}$ are mainly induced by wave moving. In case 3 and case 4, similar tilted downwind patterns of $\tilde{u}$ as that in case 1 can be observed due to the sheltering effect in the region past the wave crests. The magnitude of $\tilde{u}$ in wind fields is larger than that in the water, which indicates that the wave coherent velocity $\tilde{u}$ in case 1, case 3 and case 4 is caused by the fast moving wind above the waves. In case 4, $\tilde{u}$ is confined to a smaller region close to the wave surface because the wave height is lower. The negative and positive patterns of $\tilde{w}$ are opposite to that under wave surface, which is consistent with the results in case 1.

\begin{figure}[h!]

\centering\includegraphics[width=0.9\linewidth]{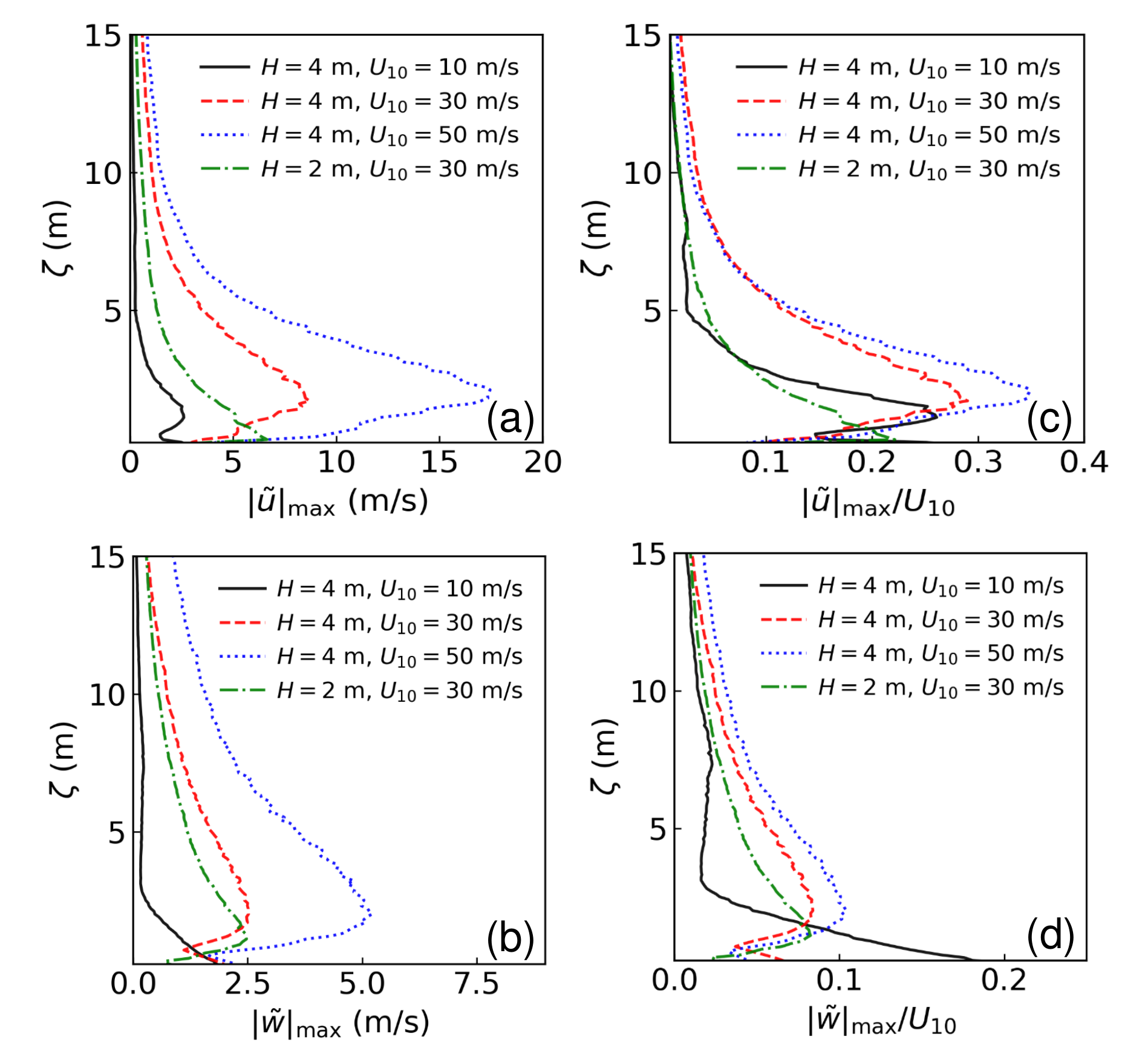}.
\caption{Vertical profiles of maximum wave coherent velocity across all phases: (a) $|\tilde{u}|_{\mathrm{max}}$; (b) $|\tilde{w}|_{\mathrm{max}}$. Vertical profiles of normalized maximum coherent velocities: (c) $|\tilde{u}|_{max}/U_{10}$; (d) $|\tilde{w}|_{max}/U_{10}$. }

\label{fig:caseStudy_waveCoherence_verticalProfile}
\end{figure}

Figs. \ref{fig:caseStudy_waveCoherence_verticalProfile}(a) and (b) illustrate the vertical profiles of the maximum coherent velocities $|\tilde{u}|_{max}$, $|\tilde{w}|_{max}$ across all phases in the four cases. Figs. \ref{fig:caseStudy_waveCoherence_verticalProfile}(c) and (d) show the normalized maximum coherent velocities $|\tilde{u}|_{max}/U_{10}$, $|\tilde{w}|_{max}/U_{10}$. From Figs. \ref{fig:caseStudy_waveCoherence_verticalProfile} (a) and (b), one can find that, over old waves, the wave coherent velocity $\tilde{u}$ and $\tilde{w}$ are intensified ($0.05U_{10}$) close to the wave surfaces up to a height of 4 m and 2 m separately. It is noted that the slightly varied $\tilde{u}$ and $\tilde{w}$ close to the wave surface are caused by the moving of waves. Over young waves, $\tilde{u}$ and $\tilde{w}$ are larger and are intensified to a higher region. The $\tilde{u}$ and $\tilde{w}$ are induced by the moving of wind. Comparing the red and blue lines in Fig. \ref{fig:caseStudy_waveCoherence_verticalProfile}(a), it can be found that higher wind forcing induces greater wave coherent velocities $\tilde{u}$ but affects the averaged streamwise velocity fields to a similar height of 8 m. Comparing the red and green lines in Fig. \ref{fig:caseStudy_waveCoherence_verticalProfile}(a), one can find that higher waves induce slightly higher $\tilde{u}$ and affect the $\tilde{u}$ to a higher region. From Fig. \ref{fig:caseStudy_waveCoherence_verticalProfile}(c), it can be observed that, over young waves, the wave coherent velocity $\tilde{u}$ is approximately proportional to the wind velocity. The influenced region of $\tilde{u}$ mainly depends on the wave height. For wave coherent velocity $\tilde{w}$, higher wind forcing also causes greater $\tilde{w}$ and affects the averaged vertical velocity fields to a higher region. However, different heights of waves induce $\tilde{w}$ with similar magnitudes, but higher waves influence $\tilde{w}$ to a higher region.

The presented results in this section indicate that the wind forcing shifts the waves forward and increases the wave peaks. The moving waves induce and enhance wind turbulences and cause the variation of averaged wind velocities with wave phases. The regions of intense turbulence and patterns of negative-positive wave coherent wind velocities depend on the relative speed between wind velocity and wave phase speed. Waves with higher heights induce wind turbulences to a higher region. The intensities of wave induced turbulences and wave coherent velocities, as well as the height of wave influenced region are affected by the wind velocity and wave heights. 

\section{Conclusion}

By nature, winds and water waves are coupled through exchanging momentum at the air-water interface, which affects the characteristics of wind fields and wave profiles. Due to the complex physics, it is challenging to accurately model the evolution of coupled wind and wave. Currently, existing research on two-phase wind-wave modeling mainly analyzed the coupling effect under uniform wind forcing without considering the inherent strong turbulences of wind above the air-water interface. This is adequate to realistically and accurately quantify the combined wind-wave loading effects on coastal and offshore structures. 

The present study develops a two-phase flow model to simulate coupled turbulent wind-wave fields using the open source program OpenFOAM. The volume of fluid (VOF) method is adopted to capture the complex water-air interface configuration. An advection equation for the VOF indicator function is introduced to track the motion of two-phase flows. The wind-wave field conditions at the inlet boundary are defined through specifying water particle velocities, turbulent wind velocities and wave surface elevation. The high-fidelity high-resolution two phase flow model is verified using experimental data from a laboratory scale testing. The validated model is then applied to analyze the coupled interaction of wind and wave on a 100 m scale especially under extreme wind and wave conditions. Based on the presented results, we can obtain the following conclusions.

\begin{enumerate}
  \item The wind-wave two-phase numerical model can well predict the coupled wind-wave fields and capture the details of wind turbulences. The simulated wave elevations of wave groups without wind forcing and under following wind agree well with experimental data. The characteristics of air flow above waves, such as the turbulent stresses and wave coherent stresses fluxes and the turbulent stresses, agree well with experimental results.
  
  \item Wind vortexes are produced after wind travels past wave crests, which induces the development of wind turbulences to a certain height related to wave height above wave surfaces. The resultant wind turbulences are contributed by the prescribed turbulences at the inlet boundary and the wave induced turbulences.
  
  \item Averaged wind velocities vary with wave profiles near the wave surface and the variation is enhanced by the inherent turbulences in the wind.
  
  \item The location of region of intense wind turbulence depends on the relative speed between wind and wave. The intense wind turbulence occurs at the upwind side of wave crest surface if the wave travels faster than the wind, or occurs at the downwind side if wind travels faster. The wave induced wind turbulence increases when the wind forcing velocity and/or wave height increases. Quantitatively, extreme wind forcing ($U_{10}$=50 m/s) can increase the maximum turbulence intensity by 17\%. A higher wave induces turbulence in a higher region above wave surface.
  
  \item Different relative speed between wind and wave can induce different positive-negative patterns of wave coherent velocities. When wave travels faster, the wave coherent velocities are mainly induced by the movement of waves and are confined very close to the air-water interface. When wind travels faster, the wave coherent velocities are caused by the fast moving wind. The wave coherent velocity $\tilde{u}$ is approximately proportional to the wind velocity and the influenced region $\tilde{u}$ mainly depends the wave heights. 
  
\end{enumerate}

In summary, the developed two-phase model for simulating turbulent wind-wave fields is validated and applied to analyze the coupling effect between wind and wave under extreme conditions. The developed model and presented results advance the understanding on realistically coupled wind-wave field characteristics and will serve as the foundation for future research on critical infrastructure systems exposed to wind-wave impacts. In the next step, the authors will apply the verified model to characterize the coupled wind-wave loading on structures.

\newpage
\section*{Acknowledgement}

This work was supported by the Louisiana State University Economic Development Assistantship and the research was conducted using high performance computing resources provided by Louisiana State University. The authors are grateful for all the support. 
\newpage
\clearpage

\newpage

\newpage
\bibliographystyle{elsarticle-num-names}
\bibliography{references.bib}

\end{document}